\documentclass[times, twoside, watermark]{zHenriquesLab-StyleBioRxiv}

\usepackage{mathptmx}
\usepackage{newtxmath}

\leadauthor{Klamser} 

\newcommand{\deff}{d_\mathrm{eff}}

\newcommand{\tauKd}{\tau_\mathrm{Kendall}}
\newcommand{\corr}{\mathrm{corr}}
\newcommand{\rmse}{\mathrm{RMSE}}
\newcommand{\logrmse}{\mathrm{logRMSE}}
\newcommand{\logcorr}{\mathrm{logcorr}}
\newcommand{\Expv}[1]{\mathrm{E}\left[#1\right]}
\newcommand{\cpc}{\mathrm{cpc}}

\begin{document}

\title{Inferring country-specific import risk of diseases from the world air transportation network}
\shorttitle{Infer import risk from the WAN}

% Use letters for affiliations, numbers to show equal authorship (if applicable) and to indicate the corresponding author

\author[a,b]{Pascal P. Klamser}
\author[a,b]{Adrian Zachariae}
\author[a,b,c,d]{Benjamin F. Maier}
\author[e,f]{Olga Baranov}
\author[a,b]{Clara Jongen}
\author[a,b]{Frank Schlosser}
\author[a,b, \Letter]{Dirk Brockmann}
\affil[a]{Department of Biology, Institute for Theoretical Biology, Humboldt‐Universität zu Berlin, 10115 Berlin, Germany}
\affil[b]{Robert Koch Institute, 13353 Berlin, Germany}
\affil[c]{DTU Compute, Technical University of Denmark, 2800 Kongens Lyngby, Denmark}
\affil[d]{Copenhagen Center for Social Data Science, University of Copenhagen, 1353 Copenhagen, Denmark}
\affil[e]{Division of Infectious Diseases and Tropical Medicine, University Hospital, LMU Munich, Munich, Germany}
\affil[f]{German Center for Infection Research (DZIF), Partner Site Munich, Munich, Germany}

\maketitle

%TC:break Abstract
%the command above serves to have a word count for the abstract
\begin{abstract}
 Disease propagation between countries strongly depends on their effective distance, a measure derived from the world air transportation network (WAN).
 It reduces the complex spreading patterns of a pandemic to a wave-like propagation from the outbreak country, i.e. a linear relationship to the arrival time of the unmitigated spread of a disease. However, in the early stage of an outbreak, what matters to countries' decision makers is knowledge about the relative risk of arrival of active cases, i.e. how likely it is that an active case that boarded at the outbreak location will arrive in their country.
 As accurate mechanistic models to estimate such risks are still lacking, we propose here the ``import risk'' model that defines an import probability by means of the effective-distance framework.
 The model assumes that airline passengers are distributed along the shortest path tree that starts at the outbreak's origin.
 In combination with a random walk, we account for all possible paths, thus inferring predominant connecting flights.
 Our model outperforms other mobility models, such as the radiation and gravity model with varying distance types, and it improves further if additional geographic information is included.
 The import risk model's precision increases for countries that are more connected within the WAN, and recovers a geographic distance-dependence that suggests a pull- rather than a push-dynamic of the distribution process.
\end {abstract}
%TC:break main
%the command above serves to have a word count for the abstract


\begin{corrauthor}
%\texttt{dirk.brockmann{@}hu-berlin.de}
dirk.brockmann\at hu-berlin.de
\end{corrauthor}

% !!!!!!!!!!!!!!!!!!!!!!!!!!!!!!!!!!!!!!!!!!!!!!!!!!!!!!!
% Intro
% !!!!!!!!!!!!!!!!!!!!!!!!!!!!!!!!!!!!!!!!!!!!!!!!!!!!!!!
\section*{Introduction}

The recent decades have seen a considerable increase in mobility:
the world-wide number of passenger cars in use increased between 2006 and 2015 on average by about 4\% each year reaching approximately 1 billion in 2015 \cite{Carlier2021}, comparable with the yearly increase in the number of containers shipped over sea \cite{OECD2023} and the number of globally scheduled air passenger even increased yearly between 2004 and 2019 by about 6\% \cite{StatistaResearchDepartment2023}.
In other words, the world gets more connected passengerwise on the small (cars) and largest (air traffic) scale, and in terms of the import and export of goods.
This increased connectivity eases the distribution of anything related to goods and people, as shown for the distribution of over 400 invasive species by agricultural imports that is best predicted by the global trade network \cite{Chapman2017}.
A prime example of unwanted side effects of well-connected regions is the potential of a pandemic, accompanied by death, economic damage and the potential stigmatization of survivors, migrants and minorities \cite{Yashadhana2021, Hays2005, Daftary2017}.
Already the first plague pandemic that started 541 in the Nile Delta of Egypt spread in 8 years across the territories (Mediterranean, Northern Europe and Near East) of 2 affected empires because of the intense commerce in the Roman Empire \cite{Hays2005}.
Nowadays, the intensified exchange reduces the time until a pandemic reaches all parts of the world to months as for the 2009 H1N1 virus that spread from Mexico in 5 months to all continents \cite{Fineberg2014, Fraser2009} or the recent COVID-19 pandemic whose variants spread within a few months across the globe \cite{Jia2020, Klamser2022, Hadfield2018, Tegally2022}.

The connection strength between world regions is only partly explained by their geographic proximity.
Instead, due to historic geopolitical relations \cite{Sacco2023, Kissinger2015} pandemics spread rather along an effective distance that is derived from the world air transportation network (WAN) \cite{Brockmann2013, Iannelli2017, Gautreau2007, Gautreau2008}, or, if applied on a smaller scale, also from other means of transportation  \cite{Nohara2022, Brockmann2013}.
According to the effective distance, region B is closer to region A if the passenger flow from A to B is larger than to other destinations.
An interesting extension is the multipath effective distance, that improves the arrival time prediction of a spreading disease by including all paths of a random walker on the WAN \cite{Iannelli2017}.
The effective distance is regularly used to analyze the impact of mobility on the spread of diseases, as for example for MERS \cite{Nah2016}, Ebola \cite{Otsuki2016}, Zika \cite{Nah2016a} and most recently COVID-19 \cite{EdsbergMollgaard2022, Nohara2022, Coelho2020, Adiga2020}.
While it allows to qualitatively estimate the arrival time of a disease, its use is very limited for the description of import events of infected passengers from a specific source to target.
However, these import events are highly relevant for political decision-makers and to enable modelling predictions.

In this work, we describe these import events via the ``import probability'' $p(B|A)$, which is equivalent to the origin-destination (OD) matrix whose element $T_{BA}$ represents the number of trips from A to B, with the difference that the probability is normalized by all trips starting in A, i.e. $p(B|A) = T_{BA}/T_{A}$. 
There exist mobility models that fit the OD matrix, i.e. a reference OD matrix is needed as for the gravity model\cite{Zipf1946, Cascetta1988, Lenormand2012, Abrahamsson1998, Barbosa2018}.
Yet, it can be extremely difficult to obtain the OD matrix and most often it is estimated by small surveys  \cite{Gomez-Gardenes2018} or alongside a census  \cite{Masucci2013}, and even for the air transportation network with a booking system the OD is only an approximation since passengers increasingly book directly at the airlines (in 2015 30\% of all Lufthansa flights were booked directly which increased to 52\% in 2018  \cite{ONeill2019}) and not via the big GDS (global distribution systems) from which most OD-estimates are derived \cite{Recchi2019, Christidis2020}.
That means to exactly compute the air transportation OD matrix, bookings of all GDSs and about 900 airlines must be purchased/estimated and combined.
Thus, models that do not rely on an existing reference OD matrix are important and those either assume an underlying decision process  without integrating traffic information as the radiation model\cite{Stouffer1940, Simini2012} or they apply a maximum entropy approach to distribute the unknown OD trips along possible routes of a known traffic network  \cite{Abrahamsson1998, DeGrange2017, Englezou2021}.
However, none of the above approaches use the effective distance with its validated link to disease propagation and none is based on a mechanistic distribution process on a traffic network.
The second point is crucial, since such a process provides us a mechanistic understanding of the observed pattern, and it enables us to study how its modification affects the passenger distribution, e.g. how a containment intervention along the distribution paths reduces the import probability of infected passenger.

In this work, we introduce the import risk model that is based on a distribution process following the effective distance shortest path tree of the WAN combined with a random walker that explores all possible paths of the WAN from 2014.
As a ground truth, we use the \textit{Global Transnational Mobility Dataset} from 2014 \cite{Recchi2019} and investigate the discrepancy to the import risk and alternative mobility models as the gravity \cite{Zipf1946, Barbosa2018} and radiation model  \cite{Simini2012} through multiple comparison measures.
We find that the import risk model outperforms the alternatives and only marginal improves if geodesic distance information is included.
Finally, we analyze the quality of import probability estimation for certain countries and assess if and how the geodesic distance is encoded in the import risk estimate.

% !!!!!!!!!!!!!!!!!!!!!!!!!!!!!!!!!!!!!!!!!!!!!!!!!!!!!!!
% The WAN, OD-probability and the effective distance
% !!!!!!!!!!!!!!!!!!!!!!!!!!!!!!!!!!!!!!!!!!!!!!!!!!!!!!!
\section*{Relating the WAN, OD-probability and the effective distance}

\begin{figure*}
    \centering
    \includegraphics[width=\textwidth]{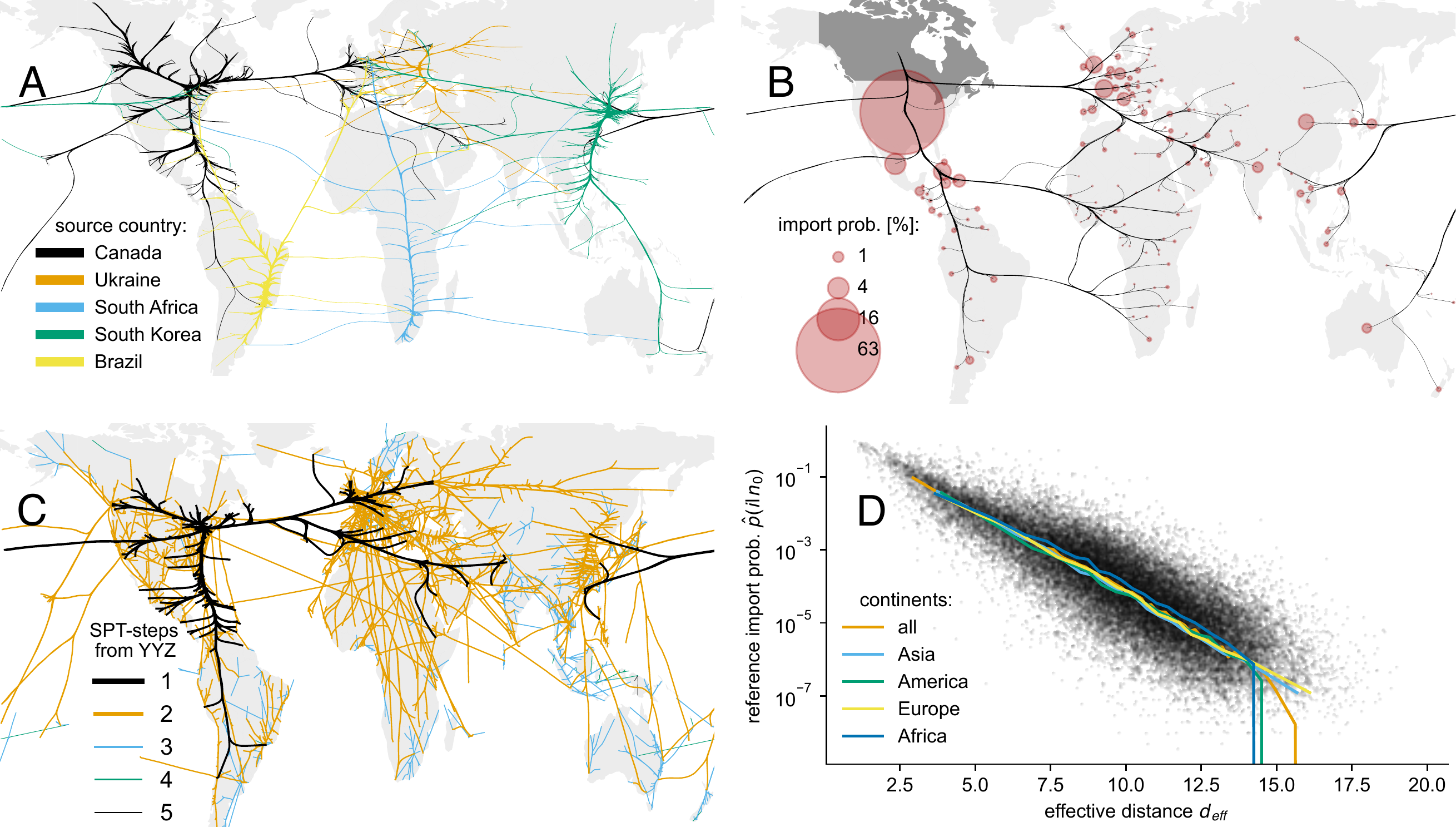}
    \caption[The relation between WAN, OD-probability, SPT and effective distance.]{
        \textbf{The relation between WAN, OD-probability, SPT and effective distance.}
        \textbf{A}: The world air transportation network (WAN) represents the direct flight connections and maximal seat capacities between airports in 2014, here shown for flights starting from five selected countries. It is based on flight-schedule-data. The lines are bundled and do not represent the specific flight route, but illustrate the links to airports abroad.
        \textbf{B}: The reference import probability from Canada to all countries, based on the OD matrix (Origin-Destination) of the \textit{Global Transnational Mobility Data set}  \cite{Recchi2019, Recchi2019_zenodo} in 2014. It combines origin and final-destination trips between countries from the SABRE and the World Tourism Organization (UNWTO). The lines illustrate the connection to the common source country.
        \textbf{C}: Based on the effective distance $\deff = d_0 - \ln(p)$ a shortest path tree (SPT) is constructed with the largest Canadian airport as source (YYZ: Toronto Pearson International Airport). The link color and thickness shows the hop distance, i.e. number of connecting flights.
        \textbf{D}: exponential decay of the reference import probability (as in \textbf{B} but for all countries as source) with the effective distance $\deff$ (derived from the SPT (\textbf{C}) of the WAN (\textbf{A})). Each dot represents a country-country link, the lines are medians including either all source countries or only from a specific continent.
    }
    \label{fig:maps}
\end{figure*}

In this work, we introduce the import risk that estimates the probability of a passenger departing in airport A to end its journey at any other airport world-wide, including airports not directly connected to the origin airport.
The estimation is based on the traffic flow of airplanes and the respective maximal passenger capacity between airports, a.k.a. the world air transportation network (WAN), provided by the Official Airline Guide (OAG) \cite{oag_schedule_data}.
This inference-problem is intriguing because it is much easier to monitor the origin and destination of airplanes, than of passengers with possibly multiple connecting flights until their final destination.  
In our study we use the WAN from 2014 (Fig.~\ref{fig:maps}A) and compare the derived import probabilities to a reference dataset.
The reference import probability is based on the \textit{Global Transnational Mobility Dataset} (GTN) from 2014 \cite{Recchi2019, Recchi2019_zenodo}, that uses a combination of an origin-final-destination dataset of a major global distribution system (GDS) and a tourism dataset from the World Tourism Organization (Fig.~\ref{fig:maps}B, see Material and Methods for more details on the data).
Before introducing the import risk model, we contrast the two datasets, introduce the effective distance \cite{Brockmann2013} and quantify its potential as the base metric for our proposed model. 

By comparing the world air transportation network (WAN) with the country-specific reference import probability from the GTN (compare Fig.~\ref{fig:maps}A, B), we see that the airports connected via direct links belong to countries that also have a high import probability.
Due to physical constraints and logistic optimization, however, not all countries that have non-zero import probability are directly linked to an airport of the source country, but are reached via connecting flights instead.
For the import probability, geodesic distance and population of the target country are useful estimates but fail in specific cases, e.g. the import to Italy is about $1.4$ times larger than to Germany even if the latter is geographically closer to Canada and has a larger population.
The effective distance is an alternative, network-based distance measure that does not rely solely on direct connections and geographic information \cite{Brockmann2013, Gautreau2008, Gautreau2007, Iannelli2017} but instead uses the passenger flow $F_{ij}$ to $i$ from $j$ and its relation to the outflow $F_j$ in form of the transition probability $P_{ij}= F_{ij} / F_j$.
Additionally, it increases the distance by a constant $d_0$ for every connecting flight:
\begin{align}\label{eq:deff}
    \deff (i|j)= d_0 - \ln(P_{ij}).
\end{align}
The effective distance between airports without direct connection is the cumulative distance along the shortest path tree (SPT) derived from $\deff$, as illustrated for the largest Canadian airport (Toronto Pearson Airport, YYZ) in Fig.~\ref{fig:maps}C.
Former studies showed that the countries' arrival time of diseases depends linearly on their effective distance   \cite{Brockmann2013, Gautreau2008, Gautreau2007, Iannelli2017}.
We show that the import probability also correlates with $\deff$ (Fig.~\ref{fig:maps}D), whereby the correlation is higher than for other distance measures (see Fig.~\ref{fig:ref_imp_vs_dists}).
In fact, the import probability decays exponentially with effective distance (linear decay on a semi-log scale in Fig.~\ref{fig:maps}D) which can be reproduced in a simplified model for a passenger that travels at a constant effective speed and has constant exit rate.
Therefore, the effective distance seems to be a good representation of the underlying distribution process, and is a promising candidate for the base of our proposed import risk model, to directly estimate the import probability.

% !!!!!!!!!!!!!!!!!!!!!!!!!!!!!!!!!!!!!!!!!!!!!!!!!!!!!!!
% Import Risk Model
% !!!!!!!!!!!!!!!!!!!!!!!!!!!!!!!!!!!!!!!!!!!!!!!!!!!!!!!
\section*{Import risk model}

\begin{figure}
    \centering
    \includegraphics[width=1\linewidth]{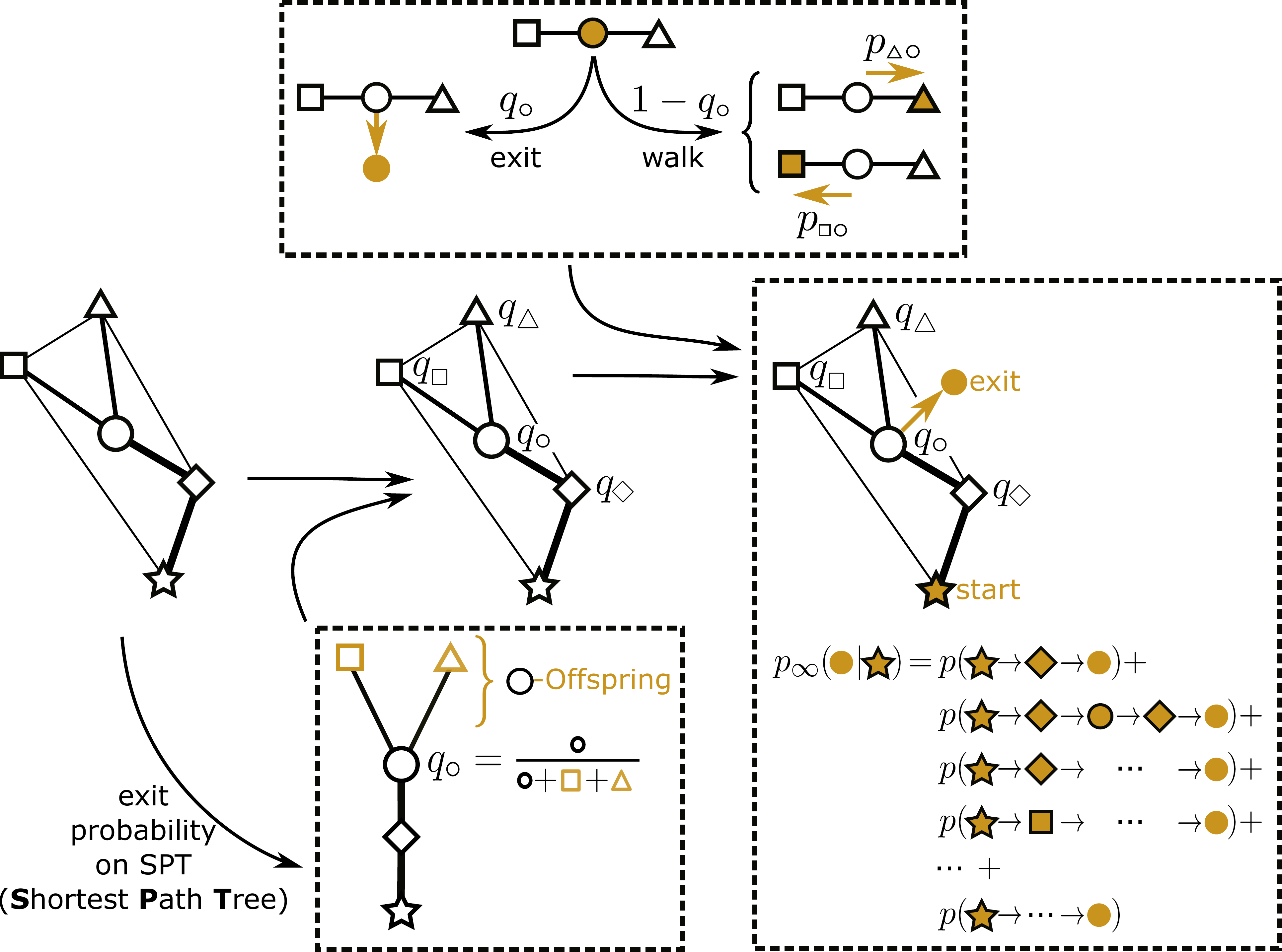}
    \caption[Import risk scheme.]{
        \textbf{Import risk scheme.}
        Starting from the transition network (left) the shortest path tree is computed based on the effective distance (center bottom).  Based on the shortest path tree, the exit probabilities $q_\circ = q(\circ|\star)$ are computed. 
        In the formula, the geometric symbols represent the estimated population of the respective node, which can also be distance-weighted (depending on the exact model).
        A random walk-process with exit probability is defined (top): at each step, the walker either exits  the node with prob. $q_\circ = q(\circ|\star)$, or continues walking with prob. $(1-q_\circ)$.
        The import risk $p_\infty(\circ|\star)$ (right) is the probability of a walker to exit at node $\circ$ given it started at node $\star$ under consideration of all possible paths.
    }
    \label{fig:method}
\end{figure}

The idea behind the import risk model is a combination of (i) a random walk with exit probability and (ii) a distribution mechanism derived from the $\deff$ SPT (Fig.~\ref{fig:method}).
Applying a random walk is motivated by \citet{Iannelli2017} who could improve the arrival-order prediction of $\deff$ by including all possible paths.
In the first step, we use the transition network representation of the WAN and let a random walker start at source $n_0$ and after each step it either exits at the current node $i$ with exit probability $q_i$ or continues to walk.
Let us define the walker's \textit{probability to continue walking} to node $n$ given it was at node $n-1$ before and originally started in $n_0$ by
\begin{align}\label{eq:S_continue}
    S_{n, n-1}(n_0) = P_{n, n-1}(1-q_{n-1}(n_0))\ ,
\end{align}
with $P_{n, n-1}$ as the transition probability from $n-1$ to $n$.
Now the probability to walk along a path $\Gamma$ starting at $n_0$ and exiting at $n$ is the probability to continue walking $S_{i, j}$ along each link $(i,j)$ that is part of the path times the exit probability of the final node
\begin{align}
    p(\Gamma) = q_{n} \prod_{(i, j)\in \Gamma} S_{i, j}\ ,
\end{align}
where we omitted the explicit dependence on the source $n_0$.
Our goal is to describe all possible paths the walker can take from $n_0$ to $n$.
We will use the matrix $\mathbf{S}$, whose elements are the probabilities to continue walking $S_{i, j}$.
The element $(i,j)$ of the product of the matrix with itself $\mathbf{S}\cdot\mathbf{S}=\mathbf{S}^2$ sums over all paths of length $l=2$ that end in $i$ and start in $j$. 
We can now specify the probability of a walker to exit at $n$ after taking all paths of length $l$ to
\begin{align}
    p_l (n|n_0) = q_n \left(\mathbf{S}^l\right)_{n, n_0}\ .
\end{align}
Finally, the import risk is the probability to exit at $n$ given all paths of all lengths
\begin{align}\label{eq:ir_final}
    p_\infty (n|n_0) &= q_n \left(\sum_{l=1}^\infty \mathbf{S}^l\right)_{n, n_0} \\ \nonumber
                     &= q_n \left( (\mathbf{I} - \mathbf{S})^{-1} - \mathbf{I} \right)_{n, n_0}\ ,
\end{align}
where we used the convergence of the geometric series with identity matrix $\mathbf{I}$.

In the second step, we approximate the exit probability $q_i(n_0)$ that we used above, but did not specify yet. 
Thereby, we assume that passengers start in source airport $n_0$, travel along the SPT and exit at node $i$ with an exit-probability
\begin{align}\label{eq:exit_prob}
    q_i(n_0) = \frac{N(i)}{N(i) + N(\Omega(i|n_0))} 
\end{align}
with $N(i)$ as the population at airport $i$ and $\Omega(i|n)$ as the set of all offspring nodes downstream of $i$ on the SPT centered at source $n_0$.
Thus, the exit probability at $i$ is the ratio of the population at $i$ to all of $i$'s downstream nodes populations, including $i$.

We approximate the population at airport $i$ with its outflow on the WAN $N(i)=F_i$ and aggregate the import probabilities on country level by summing the targets and applying a weighted average on the source airports with the population as weight.

To clarify how additional information about the geographic distance between nodes influences $p_\infty$, we explore two variations of the import risk model: In the variation with ``geodesic distance weighted'' exit probability the populations in Eq.~\ref{eq:exit_prob} are substituted with $\hat{N}(i|n_0) = N(i)/d_{i,n_0}$, where $d_{i,n_0}$ is the geodesic distance between $i$ and $n_0$.
To control for increasing model complexity, we study the ``effective distance weighted'' exit probability, where $\hat{N}(i|n_0) = N(i)/\deff(i|n_0)$, i.e. no geographic information is used, but the model structure is equivalent.

% !!!!!!!!!!!!!!!!!!!!!!!!!!!!!!!!!!!!!!!!!!!!!!!!!!!!!!!
% Alternative models
% !!!!!!!!!!!!!!!!!!!!!!!!!!!!!!!!!!!!!!!!!!!!!!!!!!!!!!!
\subsection*{Alternative models}

Many alternative models estimate the OD-matrix from which the import probability can be computed  \cite{Barbosa2018, Abrahamsson1998, Song2010, Simini2012, Brockmann2006, Stouffer1940, Schlapfer2021, Noulas2012}.
Among those, the gravity \cite{Zipf1946} and the intervening opportunity \cite{Stouffer1940, Simini2012} model are most widely used.
A recent version of the latter is the radiation model \cite{Simini2012}. It is derived from a mechanistic decision process and is in consequence parameter free, and therefore similar to our model. However, it only requires information on the population density and does not integrate flight data.
We compare our model to the gravity model with an exponential and power-law distance dependence and the radiation model (see Material and Methods for definitions).
These models only use the outflow from the WAN to estimate the node's population and the geographic locations.
To incorporate structural information of the WAN, the alternative models are also implemented with the geodesic path distance (the geodesic distance along the SPT) and the effective distance, i.e. there are in total nine alternative models: the radiation model, the gravity model with exponential and with power-law distance decaying function, and each implemented with geodesic, geodesic path and effective distance.
The exponents of the six gravity models are fitted to the reference import probability by assigning the best fitting exponent to each of the six comparison measure (Pearson correlation, root-mean-square error, common part of commuters, Kendalls rank correlation and the correlation and RMSE of the logarithmic measures, all defined in Material and Methods) and taking their mean value (see Figs.~\ref{fig:SI_gravity_scans},\ref{fig:SI_best_para_gravity}).
As comparison measures, we have chosen three measures that are related to the absolute error and three that are related to the relative error between estimate and reference.

\begin{figure*}
    \centering
    \includegraphics[width=0.9\linewidth]{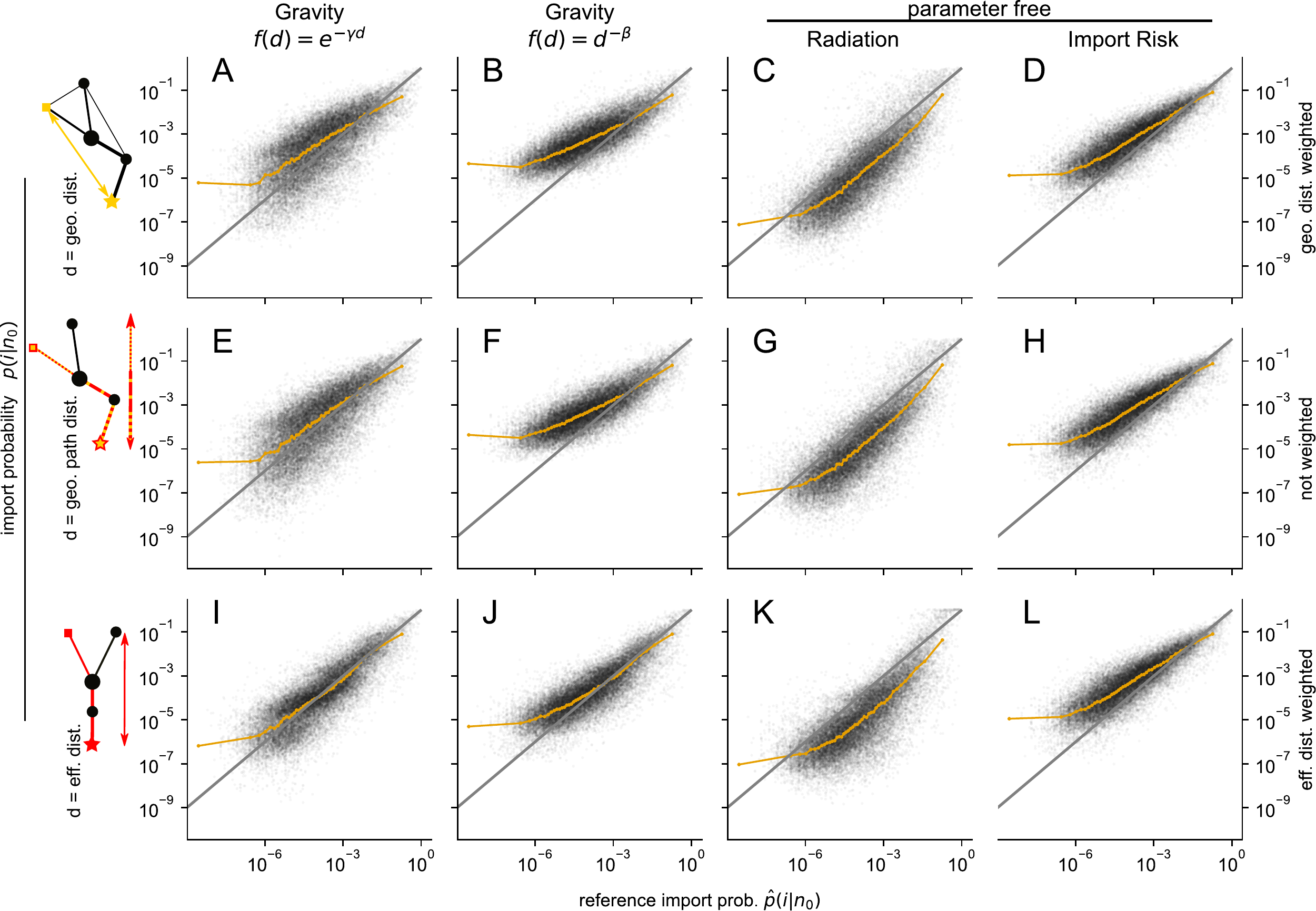}
    \caption[Model estimates of import probability]{
        \textbf{Estimates of import probability}
        by the gravity model with exponentially (1st column) and power law (2nd column) decaying distance function, the radiation model (3rd. column) and by the import risk model (4th column).
        The first three models (1st-3rd column) use as distance the geodesic (1st row), geodesic path (2nd row) and the effective (3rd row) distance.
        The import risk model is computed from the WAN with the geodesic distance (\textbf{D}) or the effective distance (\textbf{L}) as a weight for the exit probabilities or without weighting (\textbf{H}), i.e. in the last two cases (\textbf{H, D}) only WAN information is used.
        The orange line depicts the median and the gray line is $y=x$ and illustrates perfect mapping.
    }
    \label{fig:ir_vs_ref}
\end{figure*}

\subsection*{Symmetry by returning visitors}
Each of the twelve models provides an estimate for the import probability $p(i|n_0)$, from which the OD-matrix $\mathbf{T}$ can be computed by a multiplication with the respective source population $N(n_0)$.
By comparing the symmetry of $\mathbf{T}$ with the reference OD-matrix $\hat{\mathbf{T}}$, we find a much higher and qualitatively different symmetry in the reference data (see \cref{SIsec:symm}, Fig.~\ref{fig:SI_symm_check}).
The high symmetry is likely due to visitors (family, business, tourism, etc.) that dominate the international travel. They return to their home-location after a limited period \cite{Belik2011} and only the minority of the travelers are migrants, i.e. stay permanently at the destination.
Interestingly, the import risk model has the highest symmetry, but is still less symmetric than the reference data by a factor of 4.  
Thus, before comparing the estimates in detail we correct the import probability estimates by symmetrizing their OD-matrix (by taking the symmetric part and recomputing the import probability, see Material and Methods and \cref{SIsec:symm} for details).
This correction can be seen as an alternative version of a doubly constrained model where normally the constraints on in- and out-flow are ensured by an \textit{iterative proportionate fitting}  \cite{Barbosa2018}.

% !!!!!!!!!!!!!!!!!!!!!!!!!!!!!!!!!!!!!!!!!!!!!!!!!!!!!!!
% Model Comparison
% !!!!!!!!!!!!!!!!!!!!!!!!!!!!!!!!!!!!!!!!!!!!!!!!!!!!!!!
\section*{Model comparison}
In the following, we compare the import probability estimates to the reference data:(i) directly and via their medians to analyze potential systematic errors, (ii) by a collection of six different goodness of fit measures whereby we point out the rank and the relative performance of the single model and (iii) by a classification task that is highly relevant in a pandemic context when the countries at highest import risk needs to be known.

\subsection*{Qualitative Comparison}
In Fig.~\ref{fig:ir_vs_ref} the import probability estimate $p(i|n_0)$ of each model is compared to the reference import probability $\hat{p}(i|n_0)$.
The median of the probability estimates is for the gravity models in best agreement with the reference data if the effective distance is used (Fig.~\ref{fig:ir_vs_ref} first and second column), while the median values of the radiation and import risk model are less affected by the change in distance metric or the weighting by it (first and fourth column).
All models overestimate the lowest median import probability (leftmost orange dot in Fig.~\ref{fig:ir_vs_ref}), since a large portion of the observed reference import probability is zero due to a limited number of departing passengers.
The overestimation of the median import probability persists up to $p(i|n_0) \leq 10^{-4}$ for the gravity and import risk models, but not for the gravity model with exponential distance decaying function and effective distance metric (Fig.~\ref{fig:ir_vs_ref}I) which shows the best agreement of the median with the reference data.
The radiation models (first column) systematically overestimates the highest import probabilities ($p(i|n_0)\gtrapprox 10^{-1}$) and in consequence underestimates the lower import probabilities.

\subsection*{Goodness of fit by multiple measures}
\begin{figure}
    \centering
    \includegraphics[width=1\linewidth]{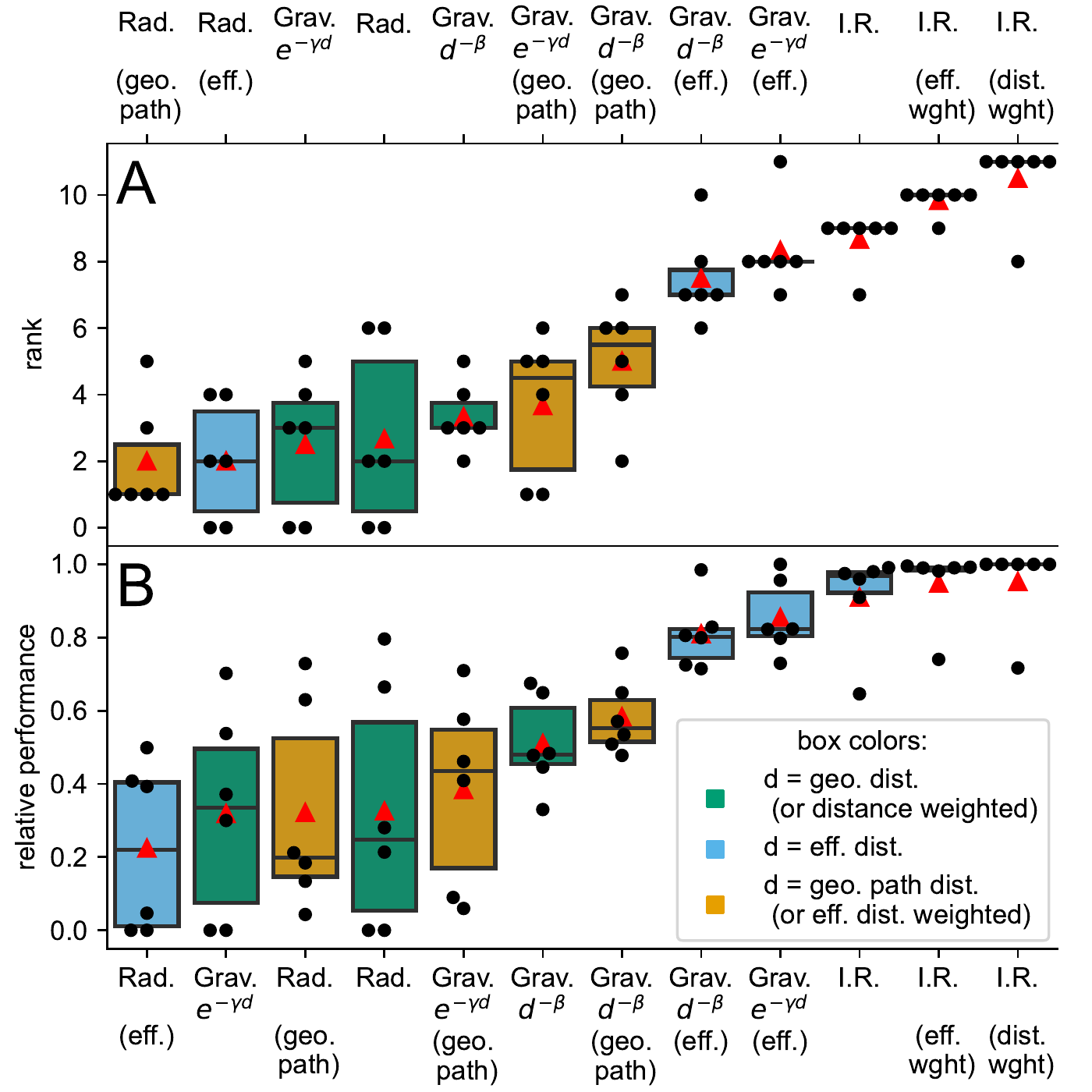}
    \caption{
        \textbf{Rank and relative performance of import risk estimation models.}
        The different import probability models are compared via their rank (\textbf{A}) and relative performance (\textbf{B}), with the highest values representing the best approach.
        The rank and relative performance are shown for each (black dots) of the six comparison measures ($\corr$, $\logcorr$, $\rmse$, $\logrmse$, $\cpc$, $\tauKd$) the box illustrates the  interquartile  range, the horizontal line the median and the red triangle the mean.
        The colors of the boxes illustrate the different distance measures in use.
        The outlier measure of the import risk models (I.R.) is the $\logrmse$, where the gravity models with effective distance are performing best.
        See Material and Methods for definitions of comparison measures and Figs.~\ref{fig:SI_model_compare_lines},~\ref{fig:SI_model_comp_absolute} for absolute and detailed relative performance.
    }\label{fig:rank_relperf}
\end{figure}

We compared each model with the reference import probability via the Pearson correlation, the root-mean-square error ($\rmse$), and the common part of commuters.
These measures are more sensitive to strong links, i.e. large import probabilities, which is important when the emphasis is placed on the countries that are most likely to import passengers. 
However, if the focus is to get a fair comparison including all links, logarithmic versions of the above measures or rank correlations are more appropriate.
Thus, we also quantify the agreement by the correlation and the RMSE of the logarithm of the measures and by Kendall's rank correlation.
The three import risk model variations outperform the other models in all but one measure, illustrated by their highest ranks, whereby the variation with geodesic distance weighted exit probability performs best (Fig.~\ref{fig:rank_relperf}A).
The import risk models are followed by the two effective-distance based gravity models, the other models don't show a consistently high rank for all 6 measures, but are rather homogeneously scattered in the lower half.
This model categorization also holds for the relative performance of the models (Fig.~\ref{fig:rank_relperf}B), with linear scaling of values in between (see Eq.~\ref{eq:rel_perf}).
% where $1$ is the best and $0$ the worst performing model and a relative performance of $0.5$ means that the model is halfway between the best and worst model (see Eq.~\ref{eq:rel_perf}).
In contrast to the ranks, the median relative performance improves considerably if the effective distance is used in the gravity models, while there is only a marginal difference in between import risk models.

The only measure where the import risk models are outperformed by the gravity models with effective distance is the $\logrmse$ (Figs.~\ref{fig:SI_model_compare_lines},~\ref{fig:SI_model_comp_absolute}).
It is expected from the gravity models' good agreement in median import probability with the reference data over wide ranges and the overestimation of low import probability by the import risk model.
This overestimation can be reduced by model-modifications that introduce parameters favoring the exit at nodes with large-populations (for details, see \cref{sec:SI_overestimation_ir} and Figs.~\ref{fig:SI_import_risk_log_fit},\ref{fig:SI_import_risk_variations}).
However, we refrain from adding complexity to the model, since its generic nature is its key aspect.

\subsection*{Classification of ten top risk countries}
\begin{SCfigure*}%[tbhp]
\centering
    \includegraphics[width=0.84\textwidth]{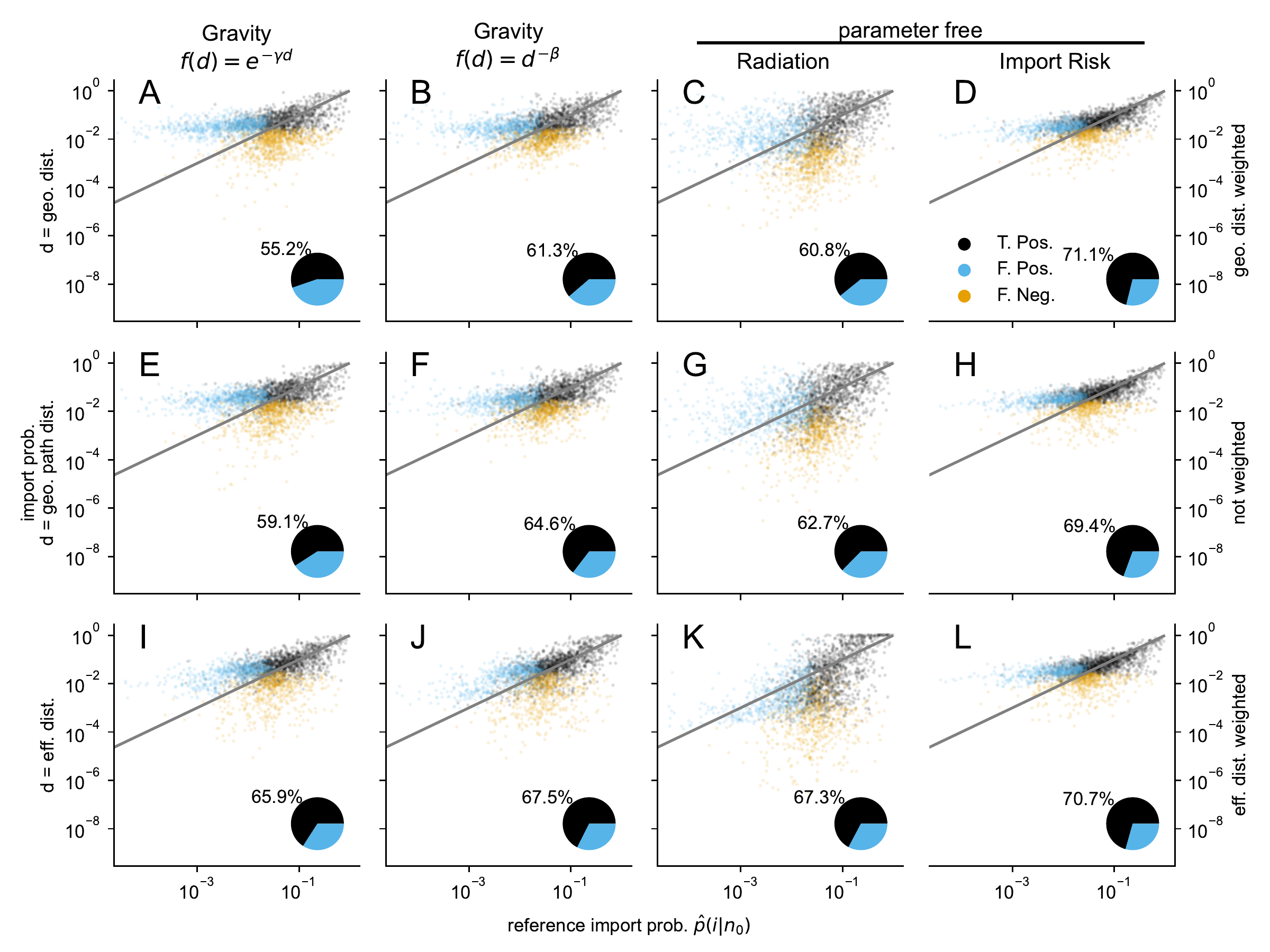}
    \caption{
    \textbf{Classification of the 10 countries with the highest import probability}
    by the gravity model with exponentially (1st column) and power law decaying (2nd column) distance function, the radiation model (3rd. column) and by the import risk model (4th column).
    A true or false positive (T. Pos. or F. Pos.) means that the country is or is not among the 10 countries with the highest reference import probability $\hat{p}$.
    A false negative (F. Neg.) means that it belongs to the reference set but was not detected by the respective model.
    The pie chart illustrates the sensitivity of the models.
    }
    \label{fig:10largest}
\end{SCfigure*}

In a pandemic context, it is of specific interest to identify the countries with the highest import probability. 
We analyzed how well the twelve proxy models can classify, if a country is among the ten countries with highest import probability.
Again, the import risk models outperform the other models and the one with geodesic distance-weighted exit probabilities is the top predictor with a sensitivity of $71.1\%$ (Fig.~\ref{fig:10largest}D).
All effective distance-based models have a high sensitivity ($\gtrsim 65\%$), including the radiation model with $66.8\%$ that had the lowest relative performance and second-lowest mean rank (Fig.~\ref{fig:10largest}I-K).
For these high import probabilities, the import risk models now outperform the other models also in terms of $\rmse$ and $\logrmse$,
i.e. the 10 countries at highest risk are not only classified best by the import risk model, but also quantitatively assessed best.

% !!!!!!!!!!!!!!!!!!!!!!!!!!!!!!!!!!!!!!!!!!!!!!!!!!!!!!!
% Import risk of countries and regions
% !!!!!!!!!!!!!!!!!!!!!!!!!!!!!!!!!!!!!!!!!!!!!!!!!!!!!!!
\section*{Import risk of countries and regions}
Having quantified the performance of the import risk model, we now focus on (i) country specific differences in its prediction quality, (ii) possible limitations due to no concept of administrative units (e.g. countries) whose airports are more interconnected   and (iii) how the geodesic distance is encoded in the import risk model, i.e. how a distance dependence emerges from WAN information only. 

\begin{figure}[t]
    \centering
    \includegraphics[width=0.98\linewidth]{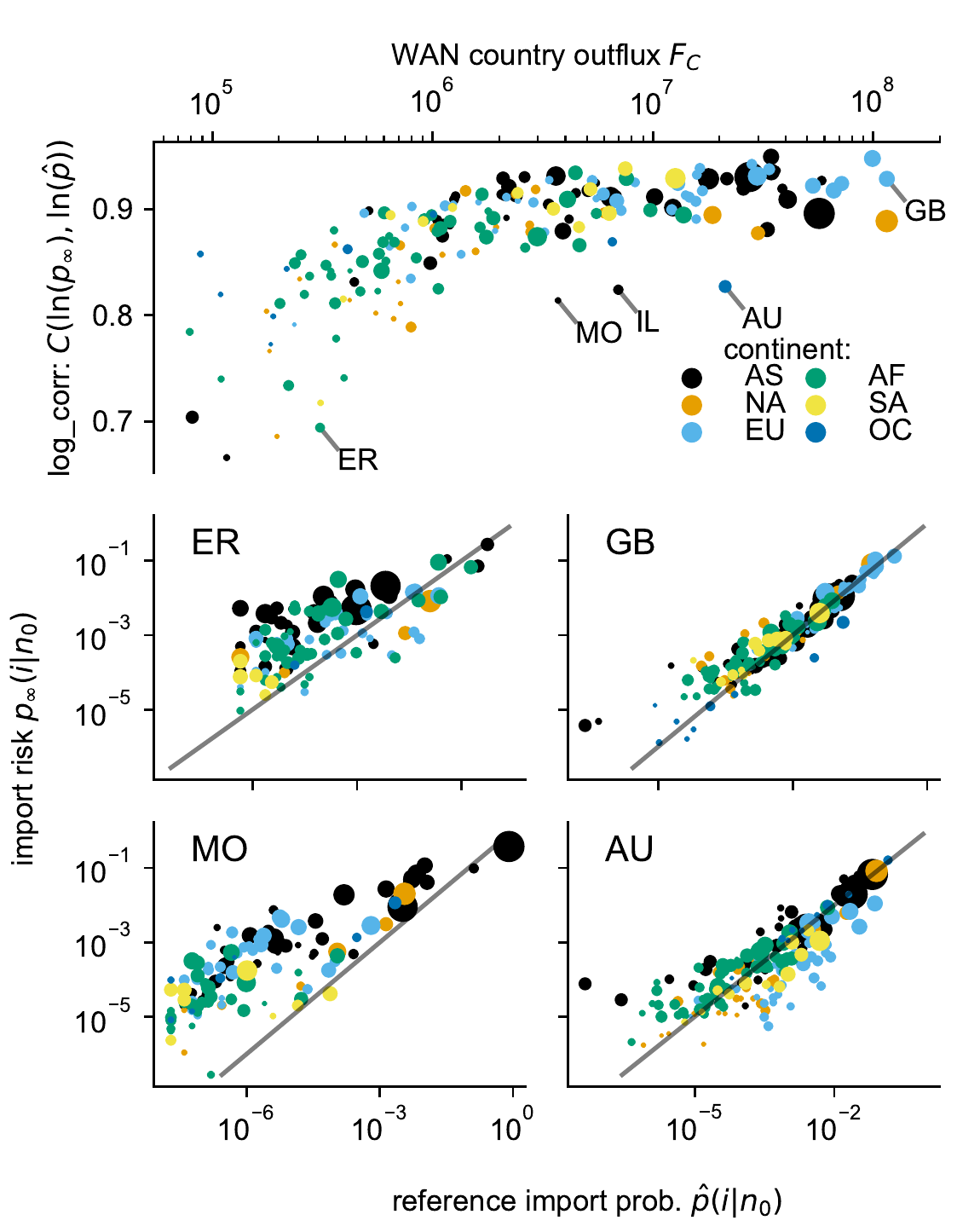}
    \caption{
        \textbf{Source countries' prediction quality and WAN outflow.}
        The correlation between the logarithm  of the import risk and the reference import probability $\logcorr = \corr(\log(p_\infty), \log(\hat{p}))$ improves with the outflow of the respective source country (top).
        Examples of source countries with particularly low (ER, Eritrea) and high (GB, Great Britain) outflow and $\log\_corr$ are shown with their import risk and reference import risk to target countries (middle row).
        Countries with exceptionally low $\log\_corr$ measures compared to source countries with a comparable outflow are either historically linked to specific regions as Australia (AU) and Israel (IL) to European countries (lower right panel) or politically as Macao (MO) as a special administrative region of China.
    }
    \label{fig:log_corr_and_examples}
\end{figure}

\subsection*{Country specific performance}
In the import risk approach, we assume minimal knowledge of the system, i.e. only the WAN is known.
Consequently, we differentiate countries only via their network properties, one of which is the degree of a node, or more precisely the node strength, since the WAN is a weighted network.
It is the simplest metric that is also easily adjustable for the country-level perspective.
For the country level the node strength directly corresponds to the flow out of country $C$
\begin{align} \label{eq:outflow_cntr}
 F_C=\sum_{n \in C} \sum_{m \not\in C} F_{mn}\ .
\end{align}
This country characteristic represents the potential of a country to dominate the structure of the network, since flows from small-outflow countries are diluted by large-outflow countries.
From an ecological point of view, the outflow is strongly correlated with the gross domestic product of a country (Fig.~\ref{fig:SI_outflow_gdp}).
The correlation ($\logcorr$) between the logarithms of the import risk $p_\infty$ and the reference import probability $\hat{p}_\infty$ improves with the outflow of the source country (Fig.~\ref{fig:log_corr_and_examples}), as illustrated by Great Britain (GB) as the country with the largest outflow in the WAN and Eritrea (ER) as one of the countries with the lowest outflow.
The prediction improvement with the country's outflow suggests that the WAN is dominated by large-outflow countries and therefore predictions worsen for countries with lower WAN outflow. 
However, the prediction improvement is also present in model alternatives that do not use WAN information at all (e.g. gravity with geodesic distance, Fig.~\ref{fig:SI_log_corr_and_examples}).
We rule the explanation out that the alternative models show this improvement due to preferential fitting of strong links -- and therefore of large-outflow countries -- since the models are fitted to the reference data by their import probabilities which ensures equal weighting among countries.
It rather suggests that the mobility behavior in low outflow regions is different, also supported by the sudden performance saturation for countries with a WAN outflow of $F_C \gtrsim 10^6$ (Figs.~\ref{fig:log_corr_and_examples},\ref{fig:SI_log_corr_and_examples}).
Possibly, their passenger distribution is constrained by additional factors and is limited to the regions in proximity.
\\

There are clear exceptions where the import risk estimation is worse compared to outbreak countries with a similar WAN outflow, as Australia (AU), Israel (IL) and Macao (MO).
These countries are connected due to historical relations to specific regions that are either not in their direct neighborhood (European countries for AU and IL) or that are more important than the bare neighborhood would suggest, as Macao that is a special administrative region of China. 
For Macao the import risk to China is underestimated which consequently overestimates the import to other countries, and for AU and IL Europe is underestimated which overestimates other regions (Fig.~\ref{fig:log_corr_and_examples}).
The exceptions AU, IL and MO illustrate that not all information is fully encoded in the WAN and therefore can not be extracted by the import risk model.
Another concept that is missing in our methodological approach is the idea of a country or another administrative unit, i.e. we treat airport pairs equally irrespective of their country-affiliation.
Since we know the international flights leaving a specific country from the WAN, we can run a self-consistency analysis, i.e. without the need of reference import probability data.
We can estimate the outflow leaving the country $C$ by the import risk model by 
\begin{align}
 T_C = \sum_{n \in C} \sum_{m \not\in C}p_\infty(m|n) N_n\ .    
\end{align}
If we compare it to $F_C$ the WAN flow out of country $C$ (see Eq.~\ref{eq:outflow_cntr}), it turns out that the import risk model systematically overestimates the flow out of a country (Fig.~\ref{fig:SI_self_consistent_outflow}A).
In fact, the relative error increases with the number of airports belonging to the country (Fig.~\ref{fig:SI_self_consistent_outflow}B).
This overestimation could be due to the lack of a country concept in the import risk model, another explanation is the overestimation of the population in the catchment area of the airport by its outflow (the outflow erroneously  counts transit passengers to the population).
However, we can easily correct for this overestimation on country-level analysis, by normalizing the airport population such that the WAN country outflow is recovered.
% note:repeat it with airpiort outflow normalized such that WAN country is recovered (then an overestimation is really due to missing country concept, but until now unclear)

% now add the part where the distance dependence is found:
% Todo: - distance dependence check also for country level, but add this only in SI (otherwise it seems strange why we shift to world-regions)
\subsection*{Geodesic distance dependence}
\begin{figure*}
    \centering
    \includegraphics[width=1\linewidth]{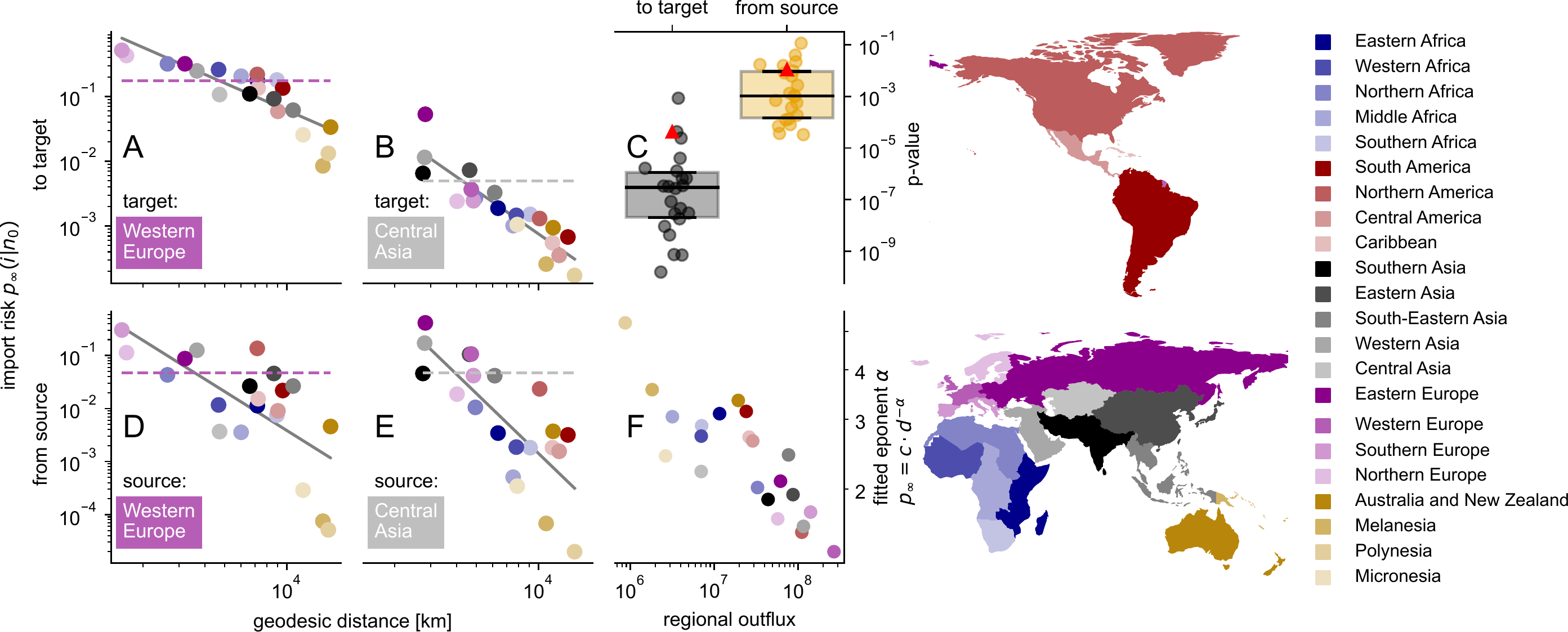}
    \caption{
        \textbf{Import risk aggregated on regional level ``to target'' vs. ``from source'' and its geodesic distance dependence.}
        The geodesic distance between regions predicts the import risk $p_\infty$ to a single target from all sources (\textbf{A}, \textbf{B}) better than from a single source to all targets (\textbf{D}, \textbf{E}) as can be seen by the p-values (\textbf{C}) of the power law fit $p_\infty(d) = c\cdot d^{-\alpha}$ that is illustrated for each selected examples by a grey line (\textbf{A}, \textbf{B}, \textbf{D}, \textbf{E}).
        The fitted exponent $\alpha$ of the import risk to a single target decreases with the respective regional WAN flow out of the target region (\textbf{F}), i.e. the more connected a region, the weaker the import risk decays with distance.
        The dashed horizontal lines  show the average import risk of a single target (\textbf{A}, \textbf{B}) or a single source (\textbf{D}, \textbf{E}).
        The color of the dots corresponds to the depicted world regions (right).
    }
    \label{fig:target_view}
\end{figure*}

The import risk model estimates import probabilities without explicit geodesic-distance information (excluding the variant with distance weighted exit probability).
Since classical models have proven distance to be a good predictor for human mobility, we assume that it is encoded in the WAN structure and by consequence in the import risk estimate  \cite{Daqing2011}.
For illustrative clarity, we aggregate the import risk on the twenty-two world regions and find that the import risks to a single target decreases with the geodesic distance to the sources in a power-law like manner (Fig.~\ref{fig:target_view}A,B and Fig.~\ref{fig:SI_ir_dist_dep_all}).
If we switch the viewpoint and analyze the distance-dependence from a single source to all target regions (Fig.~\ref{fig:target_view}D,E), the dependence is less in agreement with a power-law fit $p_\infty = c\cdot d_{ij}^{-\alpha}$ (Fig.~\ref{fig:target_view}C).
This is surprising, since the import risk is computed via a source-centric view (by computing the exit probability from the shortest path tree originating at each source), which suggest that the distance dependence should be best from one source to its possible targets.
A possible explanation is that each target has its own attractiveness independent of the source region, i.e. that the distribution dynamic resembles more a pull- than a push-dynamic.
Indeed, we find that the fitted exponent $\alpha$ from the power-law fit decreases the larger the WAN flow out of the target region is, which can serve as a proxy for the attractiveness of a region (Fig.~\ref{fig:target_view}F).
In other words, the more attractive a region, the larger the import risks from more distant source regions.
The fitted exponent $c$ has a high rank correlation with $\alpha$ ($\tauKd=0.89$), i.e. also the coefficient is dependent on the attractiveness of the region.

% !!!!!!!!!!!!!!!!!!!!!!!!!!!!!!!!!!!!!!!!!!!!!!!!!!!!!!!
% Discussion and Conclusion
% !!!!!!!!!!!!!!!!!!!!!!!!!!!!!!!!!!!!!!!!!!!!!!!!!!!!!!!
\section*{Discussion and Conclusion}

Motivated by the import probability's strong dependence on the effective distance, we implemented the import risk model based on the effective distance shortest path tree's exit probability in combination with a random walk on the WAN.
Thus, inferring the trip distribution of passengers from the traffic network of their transport vehicle (WAN).
By comparing our parameter free model with variations of established mobility models, we find that it outperforms the alternatives in all but a single comparison measure, where the two parameter-fitted gravity models with effective distance are performing best.
The import risk model is the most accurate in determining countries with the highest import probability, showcasing its importance for epidemic related problem.
However, it systematically overestimates low import probabilities and its performance worsens for countries with a passenger outflow below a million per year.
Despite the lack of any explicit geodesic distance information, the import risk model recovers a geodesic distance dependence.
This is more distinct from all sources to a single target than vice versa, which we connect to a target specific attractiveness estimated by its node strength, i.e. the target's passenger outflow.

% Reference Data:
% -modified the original data for countries that are close to each other -> if only large connections are considered results remain for most models same, only the radiation model improves significantly.
% Discussion some default import-risk settings
% - estimating the population by the outflow, alternatively use the GDP (most likely best predictor, maybe also cite some literature on that)

The only measure where the gravity models with effective distance outperform the import risk models is the $\logrmse$.
This is likely due to their good agreement over wide ranges of the import probability (Fig.~\ref{fig:ir_vs_ref}I,J).
The import risk model performs poorly with respect to the $\logrmse$ due to its systematic overestimation of low import probabilities.
Note, that the second parameter free model, the radiation model, systematically underestimates low import probabilities in the same way as import risk does.
It is expected since deviation from the assumptions can not be corrected by any parameter adjustment.
We identified several ways to reduce the import risk's overestimation of low import probabilities by introducing an additional parameter that scales the population of the respective airport, changes the exit probability along the shortest path tree or only these of the terminal nodes (for details, see \cref{sec:SI_overestimation_ir} and Figs.~\ref{fig:SI_import_risk_log_fit},\ref{fig:SI_import_risk_variations}).
We conclude that modifications that increases the probability to exit at airports/nodes with large populations reduces the overestimation.
However, we leave a possible extension of our model for future studies and highlight that it outperformed the other models in all correlation measures, illustrating its high potential.

We corrected the import probability by the symmetrization of the respective OD-matrices which corresponds to a specific form of a doubly-constrained model.
Normally, the constraints only ensure that the out- and inflow of each location corresponds to the observations \cite{Barbosa2018,Noulas2012,Lenormand2016}, in contrast, we assume that both equal each other because of returning visitors and that in consequence the OD-matrix is symmetric.
We repeated the model comparison without the correction: it reduced the agreement with the reference data for all but five of the seventy-two model-measures combinations (Fig.~\ref{fig:SI_model_comp_absolute}), which is in agreement with previous studies that report a better performance of doubly constrained models \cite{Lenormand2016}.
Importantly, the import risk model still outperforms the other models if the import probability estimates are not corrected (compare Figs.~\ref{fig:rank_relperf},~\ref{fig:SI_not_symmetrized_compare}).

We found that without providing any geodesic distance information to the import risk model, a distance dependence is recovered that is stronger for import probabilities to a single target, than from a single source, even if the import probability is computed from a source-centric view.
Since the WAN is spatially embedded and has a network dimension of three \cite{Daqing2011}, its connections reflect up to a certain degree the characteristics of the embedding space, explaining that the import risk recovers distance dependence in general.
That distance is better predicting in the target-centric view aligns well with a previous study wherein a target-specific human-mobility model can collapse mobility data to multiple targets by assigning each target a specific attractiveness that is proportional to the target's population  \cite{Schlapfer2021}.

The import risk model predictions worsen for countries with a small outflow on the WAN, and since the country's WAN outflow is proportional to its gross domestic product, the model performs less good for countries with a lower GDP, i.e. small population and/or low to middle income countries.
This is problematic, since our model infers OD information (costly in terms of monitoring) from a low-cost monitoring of traffic flow, and therefore is especially interesting for regions with limited resources.
However, we find that the model alternatives (gravity, radiation) also perform poorly for low-outflow countries and that the passenger distribution of the latter is most likely constrained by the GDP and thus limited to the target-regions in effective proximity.
To circumvent this problem, one could aggregate neighboring low-outflow countries until the conglomerate crosses the outflow threshold of $F_C = 10^6$ above which we observe a performance saturation (Figs.~\ref{fig:log_corr_and_examples},\ref{fig:SI_log_corr_and_examples}).
Of course, this compromise comes with a lower spatial resolution and we emphasize the need for future research in this direction.

% possible changes and applications
We have quantified the model performance for the world air transportation network, but it can be applied to other modes of transportation, e.g. subway system, cars, buses, trains, and future research will show if there are certain conditions that need to be fulfilled for its application.
Also, the crude estimation of the travelling population in the catchment area of an airport by the respective outflow can be improved, 
since we do not take into consideration the inflated role of hubs and the missing information about transit passengers.
The simple framework that only relies on the traffic network is appealing, but in certain scenarios its prediction can be refined by using information about the GDP, Gini-coefficient or population density.

% other alternative models
We limited our comparison to the parameter-free radiation model and the fitted gravity model, consequently missing their promising variations or other 
 alternatives \cite{Barbosa2018, Abrahamsson1998, Lenormand2016, Yang2015}.
However, the gravity model is widely applied and has been shown to perform equally well \cite{Yang2015} or better than alternatives \cite{Lenormand2016}.
There are exceptions, e.g. an iterative computation of a gravity-like model outperforms the common gravity model in  cases where the complete mobility network is not available \cite{Lenormand2012}.
Additionally, the radiation model outperforms the gravity model for long-distance connections \cite{Lenormand2016}.
Still, the simplicity of the gravity model and its adaptability by parameter adjustment make it a strong counterpart.
The model alternatives make use of the WAN-structure information by using the effective distance as done in e.g.  \citet{Ren2014} where the radiation model with time-distance was better than the travel-distance on the road network to predict the traffic on each link.
Analogously, we also found that the effective distance (that relates to the arrival time of diseases) is better than the geodesic path-distance in predicting import probabilities. 

The import risk model is fundamentally different from classic approaches that estimate OD trips from traffic data, because the latter find the OD trips that best reproduce the traffic data  \cite{Abrahamsson1998, Cascetta1988, Englezou2021, DeGrange2017}, while our model runs a distribution process on the traffic data network.
Thus, our model is mechanistic, while the classic approaches either fit and require the knowledge of the reference trip data \cite{Abrahamsson1998, Cascetta1988} or are based on the assumption that the trip distribution across the links follows the maximum entropy principle, i.e. the OD trips are considered as most likely that can be realized by the largest number of microstates \cite{Englezou2021, DeGrange2017}.
Note that maximum entropy approaches require an estimation of routes and their alternatives between each OD pair, while we allow all routes to be taken by the random walker.
Our model is in consequence -- to our best knowledge -- unique in its mechanistic nature, since it allows studying the modifications of it's underlying distribution process, for example in the form of a containment strategy meant to slow or restrict a pandemic.
A straight forward implementation could be the test of a fraction of passengers $C_i\leq 1$ at every transit airport $i$, which corresponds to reducing the probability to continue walking of an infected passenger (Eq.~\ref{eq:S_continue}) to 
\begin{align*}
    \tilde{S}_{n, n-1}(n_0, \mathbf{C}) &= (1 - C_{n-1}) \times P_{n, n-1} (1-q_{n-1}(n_0)) \ .
\end{align*}
With $\mathbf{C}=[C_1, C_2, \dots]$ one would allow for a varying testing capacity between the airports.

% !!!!!!!!!!!!!!!!!!!!!!!!!!!!!!!!!!!!!!!!!!!!!!!!!!!!!!!
% Material And Methods
% !!!!!!!!!!!!!!!!!!!!!!!!!!!!!!!!!!!!!!!!!!!!!!!!!!!!!!!
\begin{matmethods}

\subsection*{Data sources}
The WAN provided by OAG (Official Airline Guide)  \cite{oag_schedule_data} contains the number of flights and the respective maximum seat capacity $F_{i,j}$ between airports $i$ and $j$ aggregated for the year 2014.
The reference import probability $\hat{p}(m|n) = \hat{T}_{mn} / \hat{T}_{n}$ is based on the ``Global Transnational Mobility Dataset''  \cite{Recchi2019, Recchi2019_zenodo} that assigns the number of trips in 2014 $\hat{T}_{mn}$ from country $n$ to $m$ worldwide by combining the world air transportation origin-final-destination data set from the company SABRE, and cross-boarder visits with an overnight stay from the UNWTO (World Tourism Organization).
Thus, $\hat{p}(m|n)$ not only represents the mobility via air travel but also via other means (see, road, rail).
However, air travel dominates long distance trips which makes it a fair reference set of the air transportation origin-final-destination matrix.
For details on how the data sets were combined, see \cref{note:wod_data}.

\subsection*{Alternative models}
The \textbf{gravity model} states that the number of trips between regions $n$ and $m$ increase with their population sizes ($N_n$ and $N_m$) and decrease with distance $d_{nm}$
\begin{align}\label{eq:gravity}
    T_{mn} = O_n \frac{N_n \ N_m}{f(d_{nm})}\ ,
\end{align}
with $f(d)$ as a function that grows monotonically with distance $d$, most often chosen as either a power-law $f(d)=d^\gamma$ or an exponential $f(d_{nm})=e^{\gamma d}$.

In the \textbf{radiation model}, the trips from $n$ to $m$ depend on their respective population sizes $N_n$, $N_m$ (or other measures as job opportunities) and on the number of people $s_{mn}$ that are in a circle with radius $r_{mn}$ centered around location $n$ including $N_n$ and $N_m$:
\begin{align}\label{eq:radiation}
    T_{mn} = O_n \frac{N_n \ N_m}{(s_{mn} - N_m)s_{mn}}\ .
\end{align}
The import probability of both models is computed by normalizing the trips with respect to the source-region
\begin{align}
    p(m|n) = \frac{T_{mn}}{\sum_j T_{jn}} = \frac{T_{mn}}{T_{n}}\ .
\end{align}

\subsection*{Trip-symmetrization}
We correct the import probability via symmetrizing the OD-matrix by
(i) compute the estimated OD-matrix 
\begin{align}
    T^{(0)}_{m,n} = p^{(0)}(m|n) N_n
\end{align}
from the import probability estimate,
(ii) correct it by computing its symmetric part 
\begin{align}
    \mathbf{S} = (\mathbf{T} + \mathbf{T}^\top) / 2
\end{align}
and 
(iii) compute the corresponding corrected import probability via 
\begin{align}
    p^{(1)} (A|B)= \mathbf{S}_{AB} / S_B\ .
\end{align}
By going through these steps, the asymmetry is reduced heavily but still persists.
Thus, we repeat steps (i) till (iii) until $p^{(3)}(A|B)$, which returns for all models a comparable asymmetry in mean and median to the reference data (see \cref{SIsec:symm} for details).

\subsection*{Comparison measures}
We compare the import probability models with the reference data via the
Pearson correlation
\begin{align}
\corr(x, y) = \frac{\Expv{ (x - \bar{x} ) ( y - \bar{y}) }}
{\sigma_x \sigma_y}\ ,
\end{align}
with $\Expv x\equiv\bar{x}$ as average, the root-mean-square error
\begin{align}
\rmse(x, y) = \sqrt{\Expv{ (x-y)^2 }}\ ,
\end{align}
the common part of commuters \cite{Yang2015}
\begin{align}
\cpc(x, y) = \frac{ 2 \sum_{ij}\min(x_{ij}, y_{ij})}{\sum_{ij} x_{ij} + \sum_{ij} y_{ij}}\ ,
\end{align}
which is 1 if all links are identical and 0 if none of them agrees.
All the above measures are more sensitive to strong links, i.e. large import probabilities.
However, if the focus is to get a fair comparison including all links, we are more interested in logarithmic versions of the above measures or rank correlations.
Thus, we compare the logarithm of the import probabilities via correlation
\begin{align}
    \logcorr(x, y) = \corr(\log(x), \log(y))\ ,
\end{align}
root-mean-square error
\begin{align}
    \logrmse(x, y) = \rmse(\log(x), \log(y))\ ,
\end{align}
and use the Kendall rank correlation coefficient
\begin{align}
    \tauKd = \frac{C - D}{\sqrt{(C + D + T_x)  (C + D + T_y)}}\ ,
\end{align}
with $C$ and $D$ as the number of concordant and discordant pairs and $T_x$ and $T_y$ as ties only in $x$ and $y$, respectively.

To simplify and generalize the comparison we combine the six above defined measures by computing the mean rank of each model, i.e. the best correlating model has the highest (12) and the worst the lowest (0) rank and the mean rank of one model is the average of all six ranks.

To quantify the mean difference between the models we define the relative performance of one model $M$ as
\begin{align}\label{eq:rel_perf}
    \mathrm{rel. perf.}(f(x_M, y)) = \frac{f(x_M) - \mathrm{worst}(f(x_k), k)}{\mathrm{best}(f(x_k), k) - \mathrm{worst}(f(x_k), k)}\ ,
\end{align}
with $f(x_M)= f(x_M, y)$ as the specific comparison function and $\mathrm{best}(f(x_k), k)$ and $\mathrm{worst}(f(x_k), k)$ as the best and worst performing value of all models using this comparison function. Note, that $\mathrm{best}(\dots)=\max(\dots)$ apart for the rmse-measures, where it is $\min(\dots)$ (analog for $\mathrm{worst}(\dots)$).

\end{matmethods}

\begin{dataavailability}
The software to compute the import risk is available under the Zenodo repository \href{https://doi.org/10.5281/zenodo.7852476}{ImportRisk-v1.0.0} \cite{Maier2023}.

% Both the data and analysis material will be available online at Zenodo upon publication. This work is licensed under a Creative Commons Attribution 4.0 International (CC BY 4.0) license, which permits unrestricted use, distribution, and reproduction in any medium, provided the original work is properly cited. To view a copy of this license, visit \url{https://creativecommons.org/licenses/by/4.0/}. This license does not apply to figures/photos/artwork or other content included in the article that is credited to a third party; obtain authorization from the rights holder before using such material.
\end{dataavailability}

\begin{acknowledgements}
We acknowledge our families, friends, colleagues and especially Marc Wiedermann for insightful comments.
B.F.M received funding through Grant CF20-0044, HOPE: How Democracies Cope with Covid-19, from the Carlsberg Foundation and was supported as an \emph{Add-On Fellow for Interdisciplinary Life Science} while working on this study.
\end{acknowledgements}

% \section*{Bibliography}
% \bibliography{biblio}

%TC:ignore
%the command above ignores this section for word count
\onecolumn
\newpage

% \section*{Word Counts}
% \detailtexcount
% \newpage

%%%%%%%%%%%%%%%%%%%%%%%%%%%%%
% Supplementary Information HEADER %
%%%%%%%%%%%%%%%%%%%%%%%%%%%%%

\setcounter{page}{1}
\begin{center}
{\LARGE
Supplementary Information
\\
\bigskip
``Inferring country-specific import risk of diseases from the world air transportation network''
}
\\
\bigskip

{\large
Pascal P. Klamser$^{1,2}$,
Adrian Zachariae$^{1,2}$,
Benjamin F. Maier$^{1,2,3,4}$,
Olga Baranov$^{5,6}$,
Clara Jongen$^{1,2}$,
Frank Schlosser$^{1,2}$,
Dirk Brockmann$^{1,2,\ast}$,
}\\
\bigskip
$^{1}$Department of Biology, Institute for Theoretical Biology, Humboldt‐Universität zu Berlin, 10115 Berlin, Germany\\
$^{2}$Robert Koch Institute, 13353 Berlin, Germany\\
$^{3}$DTU Compute, Technical University of Denmark, 2800 Kongens Lyngby, Denmark\\
$^{4}$Copenhagen Center for Social Data Science, University of Copenhagen, 1353 Copenhagen, Denmark\\
$^{5}$Division of Infectious Diseases and Tropical Medicine, University Hospital, LMU Munich, Munich, Germany\\
$^{6}$German Center for Infection Research (DZIF), Partner Site Munich, Munich, Germany\\
$^\ast$Corresponding author E-mail:  dirk.brockmann\at hu-berlin.de
\\
\bigskip
\bigskip
\end{center}

%%%%%%%%%%%%%%%%%%%%%%%%%%%%%
% Supplementary Information %
%%%%%%%%%%%%%%%%%%%%%%%%%%%%%
\captionsetup*{format=largeformat}

% \numberwithin{equation}{section}
\renewcommand{\theequation}{S\arabic{equation}}
% renew equation-labeling:
\renewcommand{\thetable}{S\arabic{table}}
% renew equation-labeling:
\renewcommand\thefigure{S\arabic{figure}}    
\setcounter{equation}{0} 
\setcounter{section}{0} 
\setcounter{figure}{0} 

\section{Origin-destination data (``Global Transnational Mobility Dataset'')}\label{note:wod_data}

We use the ``Global Transnational Mobility Dataset''  \cite{Recchi2019, Recchi2019_zenodo} as a reference data set of the import probabilities.
It is a combination of the World-Air-Transportation-Origin-Destination (WOD) data set from the company SABRE, and cross-boarder visits (CBV) from the UNWTO (World Tourism Organization).
The WOD has in contrast to the WAN the real number of passengers from their origin airport to their final destination that booked their tickets via SABREs global distribution system (GDS).
The WTO data is based on cross border visits that include an overnight stay of non-residents, thus the backflow of residents in the country is not monitored.
The study  \cite{Recchi2019} processed and combined the two data sets by:
\begin{enumerate}
    \item decompose $\mathrm{WOD}$ in trend-, seasonal- and noise-component and only use the \textbf{trend} component timeseries :
        \begin{align*}
            T_{\mathrm{WODall}, ij} =  T_{\mathrm{WOD}, ij} &+ T_{\mathrm{WODseason}, ij}\\
                                                                 &+ T_{\mathrm{WODnoise}, ij}
        \end{align*}
    \item symmetrize the tourism flow matrix (to account for returning residents):
        \begin{align*}
            \hat{T}_{\mathrm{WTO}, ij} = T_{\mathrm{WTO}, ij} + T_{\mathrm{WTO}, ji}
        \end{align*}
    \item correct the $\mathrm{WOD}$ data since it underestimates the mobility flow for close countries (the mobility on land or water is missing):
        \begin{align*}
            \hat{T}_{\mathrm{WOD}, ij} = \left( \frac{d(i,j)}{d_{max}} \right)^{1/c} T_{\mathrm{WOD}, ij}
        \end{align*}
        with $c \approx 6.8$, $d(i,j)$ as the distance between countries $i$ and $j$ and $d_{max}$ as the maximal distance between all countries. I.e., the closer two countries, the stronger the correction and the connection of the two farthest countries is not corrected. 
    \item combine the 2 data sets by the following rules: if only one data sets provides info on the connection, take this one, otherwise take the larger flow
\end{enumerate}

\begin{align}
    \hat{T}_{ij} =
    \begin{cases}
       \hat{T}_{\mathrm{WTO}, ij} & \mathrm{if }\ \hat{T}_{\mathrm{WOD}, ij} = \emptyset \\
       \hat{T}_{\mathrm{WOD}, ij} & \mathrm{if }\ \hat{T}_{\mathrm{WTO}, ij} = \emptyset \\
       \max(\hat{T}_{\mathrm{WOD}, ij},\ \hat{T}_{\mathrm{WTO}, ij}) & \mathrm{otherwise}
    \end{cases}
\end{align}
Note that the reference data set $\hat{T}_{ij}$ is possibly an overestimation because the WOD data is increased for short connections. 
The origin to final-destination data from SABRE is derived from bookings via its GDS.
However, SABRE only had about 31\% market share in 2014 of all GDS` (Global Distribution System)  \cite{Technavio2015} and an increasing number of bookings were not done via GDS but directly via the airline company (e.g. about 30\% of all Lufthansa flights were booked directly with an increasing trend  \cite{ONeill2019}).
The WTO data is limited to overnight stays, i.e. private accommodations are not captured, also tending to underestimate the passenger flow, especially for long-range connections where passenger transport is dominated by airplanes.
Thus, the reference is only an approximation and likely underestimates the real number of passengers.

\section{Symmetrized flows}\label{SIsec:symm}

We assume, that the observed system is in equilibrium, i.e. there is no population change due to the human mobility on the WAN.
In other words, we neglect migration and assume that every visitor returns to its origin  \cite{Belik2011}, i.e. the OD-flow $T_A$ out of region $A$ is the sum of the native population $\hat{N}_A$ and the visiting populations:
\begin{align}
    O_A = \hat{N}_A + \sum_{B\neq A} \hat{N}_B p_n(A|B)
\end{align}
with $p_n(A|B)$ as the import probability of only the native population.
As a consequence, we expect the true OD-matrix to be symmetric, if it describes the human-mobility over a long time period.
The shorter the time period represented by the OD-matrix, the higher it is influenced by fluctuations (e.g. not yet all visitors returned to their origin). 
Thus, we expect a larger asymmetry between distant countries (weakly connected countries), because the few visitors might stay longer.

\begin{figure*}
    \centering
    \includegraphics[width=1\linewidth]{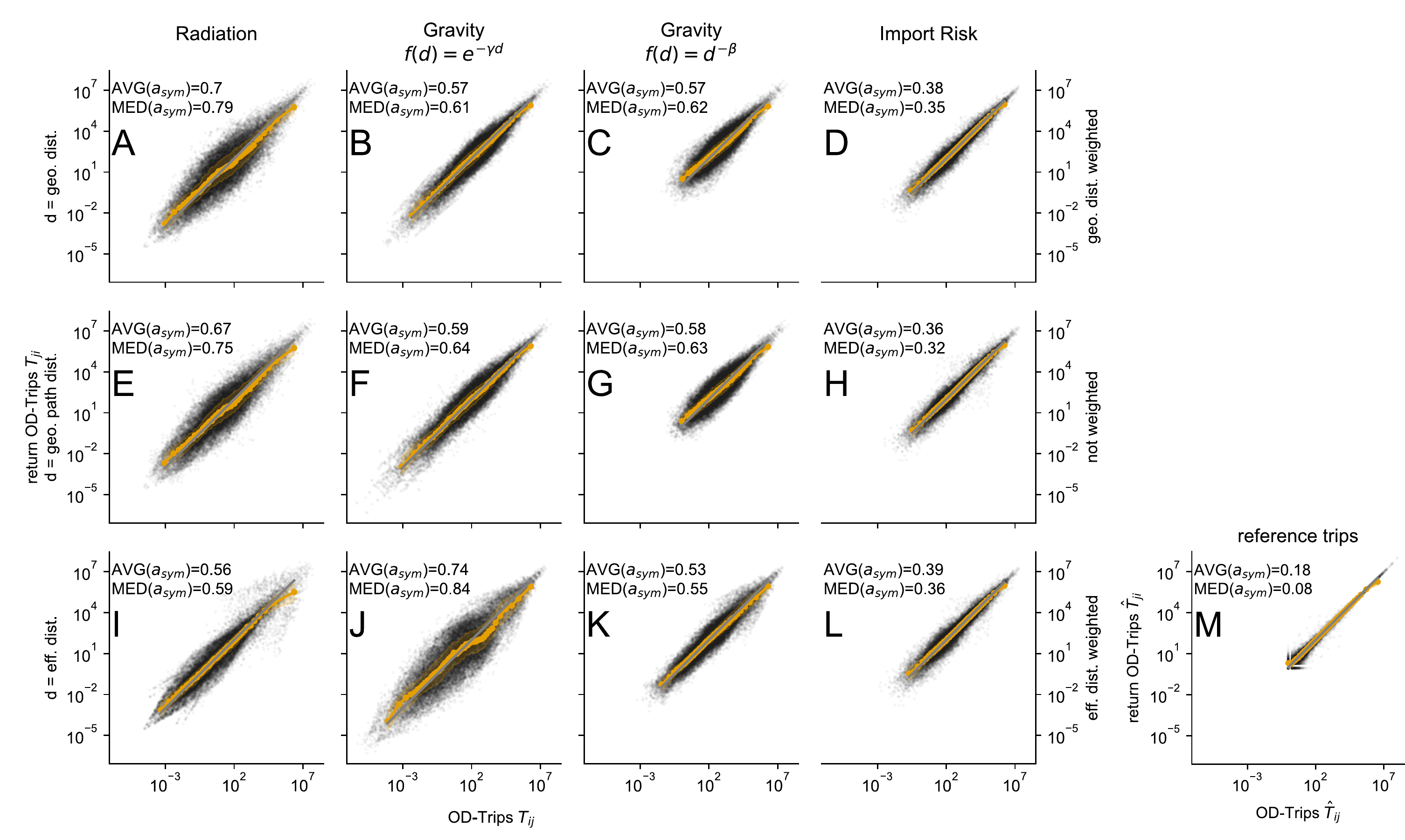}
    \caption{
        \textbf{Symmetry check for OD-matrix.}
        Each dot represents the number of passengers that travel between 2 countries and back.
        The OD-matrix is computed by the radiation model (1st. column), gravity model with exponentially (2nd column) and power law decaying (3rd column) distance function and by the import risk model (4th column).
        The OD-matrix of the models is computed by multiplying the import probability with the source-outflow. 
        The reference trips and return trips have the highest symmetry (5th column, \textbf{M}). 
        The orange line depicts the median and the gray line is $y=x$ and illustrates perfect symmetry.
        The mean (AVG($a_{sym}$)) and median (MED($a_{sym}$)) asymmetry of the flows, computed according to Eq.~\ref{eq:SI_asym}, are shown in each panel.
        The reference trips (\textbf{M}) show the lowest asymmetry, especially for large passenger flows.
        }
    \label{fig:SI_symm_check}
\end{figure*}

We estimate the asymmetry by
\begin{align}\label{eq:SI_asym}
    a_{sym}(A, B) = \frac{|T_{AB} - T_{BA}|}{\max(T_{AB}, T_{BA})}
\end{align}
and observe for the reference OD-matrix the lowest asymmetry in mean and median compared to all others estimates (compare Fig.~\ref{fig:SI_symm_check}M with A-L).
The OD-matrices estimated by the import risk model (Fig.~\ref{fig:SI_symm_check}D,H,L) are the most symmetric ones; however, the asymmetry is still twice as high than for the reference trip   (compare Fig.~\ref{fig:SI_symm_check}L with M).
Additionally, the typical pattern of lower asymmetry for stronger connections is much less pronounced in the model estimates compared to the reference data.

We symmetrize the OD-matrix by  (i) compute the estimated OD-matrix
\begin{align}
    T^{(0)}_{m,n} = p^{(0)}(m|n) N_n   
\end{align}
from the import probability estimate,
(ii) correct it by computing its symmetric part, 
\begin{align}
    \mathbf{S} = (\mathbf{T} + \mathbf{T}^\top) / 2   
\end{align}
and 
(iii) compute the corresponding corrected import probability via
\begin{align}
p^{(1)} (A|B)= \mathbf{S}_{AB} / S_B\ .
\end{align}
By going through these steps, import probabilities that represent a flow which is larger than the respective return flow are decreased and vice versa.
However, the import probability after the first correction $p^{(1)}(A|B)$ results in a OD-matrix $\mathbf{T}^{(1)}$ of higher symmetry but with still a significant asymmetry, e.g. $p^{(1)}(A|B)$ for the gravity model with exponential distance decay and effective distance the median asymmetry decreases from $MED(a_{sym})=0.84$ to $MED(a_{sym})=0.15$.
Thus, we recursively iterate $M=3$ times through steps (i) till (iii), i.e. until $p^{(3)}(A|B)$, which returns for all models a comparable $a_{sym}$ in mean and median to the reference data.

\subsection{Alternative symmetrization}
A possible reason for the asymmetry in the estimated ODs is that the import probability $p(A|B)$ only represent the import of the native population $p_n(A|B)$ from $B$ but not the visitors from $A$ returning from $B$.
However, we are interested in the combined import probability $p_c(A|B)$ that we could in principle compute by 
\begin{align}
    p_c(A|B)   &= \frac{T_{c, AB}}{T_{c, B}}\\
                   &= \frac{T_{n, AB} + T_{n, BA}}{T_{c, B}}\\
                   & =\frac{p_n(A|B) \hat N_B + T_{n, BA}}{T_{c, B}}\ .
\end{align}

The problem is, that we only know the WAN flow network $\mathbf{F}$ and thus the outflow per node 
\begin{align}\label{eq:SI_outflow_compo}
    F_A = T_A + H_A
\end{align}
which is the OD-outflow and the transit passenger $H_A$ through $A$.
We could compute the combined import probability $p_c(A|B)$, if we assume that our estimated import probability is the one of the native population, i.e. $p_n(A|B)$, and if we could estimate the native populations $\hat N_B$.
To do so, we assume that there are no transit passengers, i.e. setting in $H_A=0$ in Eq.~\ref{eq:SI_outflow_compo} we arrive at
\begin{align}
    \hat N_A = k_A F_A \quad \mathrm{with } 0 < k_A < 1\ .
\end{align}

Thus, it boils down to finding the coefficient vector $\mathbf{k}$ that estimates the combined OD-outflow 
\begin{alignat}{2}\label{eq:SI_observed_ODflux}
    \tilde T_{c, B}(\mathbf{k}) &= \sum_{A\neq B} \tilde T_{c, AB}      &&  \\
                                  &= \sum_{A\neq B} (\ \tilde T_{n, AB}  &&+ \tilde T_{n, BA}\ )  \\
                                  &= \sum_{A\neq B} (\ p_n(A|B)\ k_B\ F_B &&+ p_n(B|A)\ k_A\ F_A\ )
\end{alignat}
best compared to the true combined OD-outflow $T_{c, B} \approx F_B$.
Note that we use $\tilde{x}$ to mark estimates based on $\mathbf{k}$.
It is a high dimensional optimization problem with a bounded parameter-search space.
We used a simple square error function:
\begin{align}\label{eq:opti_error}
    e(\tilde{\mathbf{T}}_c(\mathbf{k}), \mathbf{T}_c &= \sum_A (\tilde T_{c, A}(\mathbf{k}) - T_{c, A})^2
\end{align}

We tested available optimizers for a bounded search space from \textit{scipy} (version=1.7.1, using 'scipy.optimize.minimize') by staying as close to the data as possible, i.e. using the observed outflow as native outflow and the estimated import probability as import probability of natives.
The optimizer  'Powell' and 'trust-constr' performed well on various import probability estimates, while 'L-BFGS-B', 'Nelder-Mead', 'TNC' and 'SLSQP' did not converge to the correct solution.
However, if applied to the real data, the optimizer fails for 8 of the 12 import probability estimates, i.e. the coefficients do not result in a symmetric OD-matrix (not shown).
This suggests that the assumptions do not hold, i.e. the estimated import probability does not correspond to the import probability of natives and/or transit passengers can not be neglected. 

% !!!!!!!!!!!!!!!!!!!!!!!!!!!!!!!!!!!!!!!!!!!!!!!!
% ON THE OVERESTIMATION OF LOW IMPORT PROBABILITIES
% !!!!!!!!!!!!!!!!!!!!!!!!!!!!!!!!!!!!!!!!!!!!!!!!
\section{On the overestimation of low import probabilities}\label{sec:SI_overestimation_ir}

\begin{SCfigure*}
    \centering
    \includegraphics[width=0.7\linewidth]{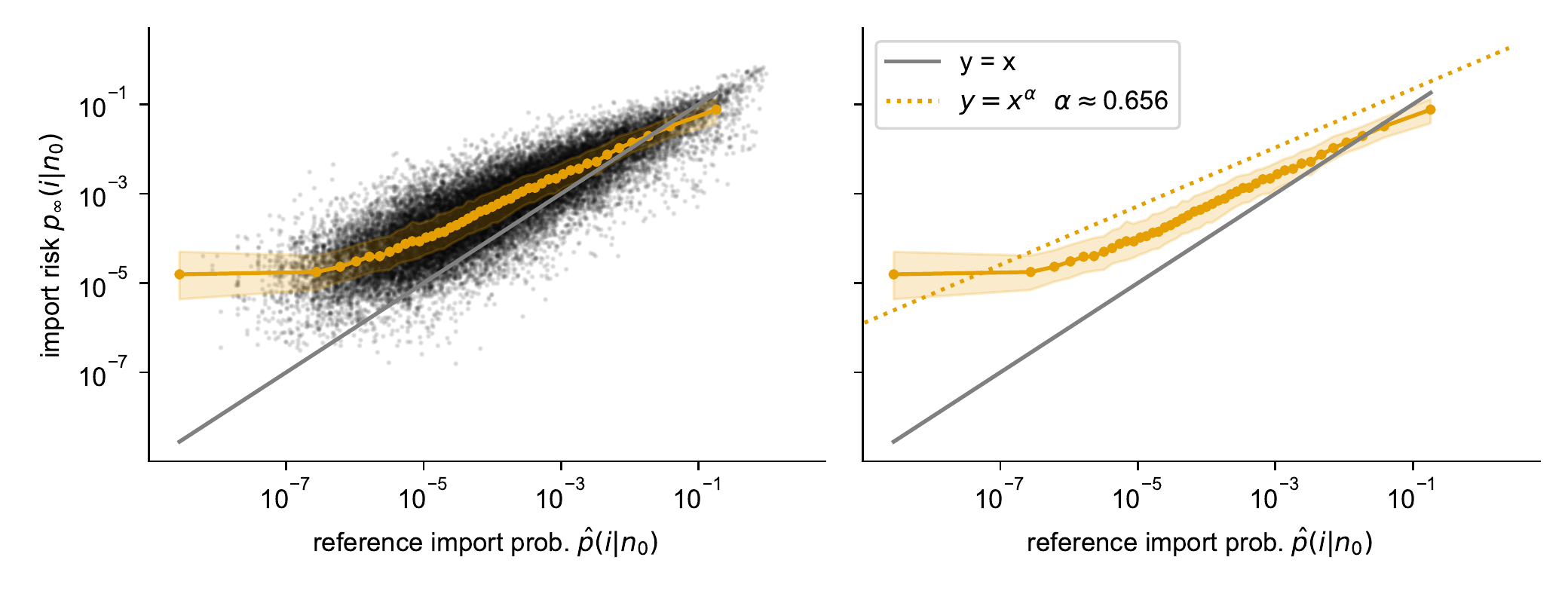}
    \caption{
        \textbf{Import risk comparison and its deviation from a linear relation.} 
        Scatter plot (left) and only median and IQR with an exponential fit (right).
    }
    \label{fig:SI_import_risk_log_fit}
\end{SCfigure*}

The import risk model does overestimate low import probabilities (Figs.~\ref{fig:SI_import_risk_log_fit}), i.e.
\begin{align}
    p_{\infty} \propto \hat{p}^\alpha,\ \alpha \approx 2/3\ . 
\end{align}
Here we present several attempts to understand mechanistically why this overestimation happens and introduce slight variations of the import risk model.

\begin{figure}
    \centering
    \includegraphics[width=0.8\linewidth]{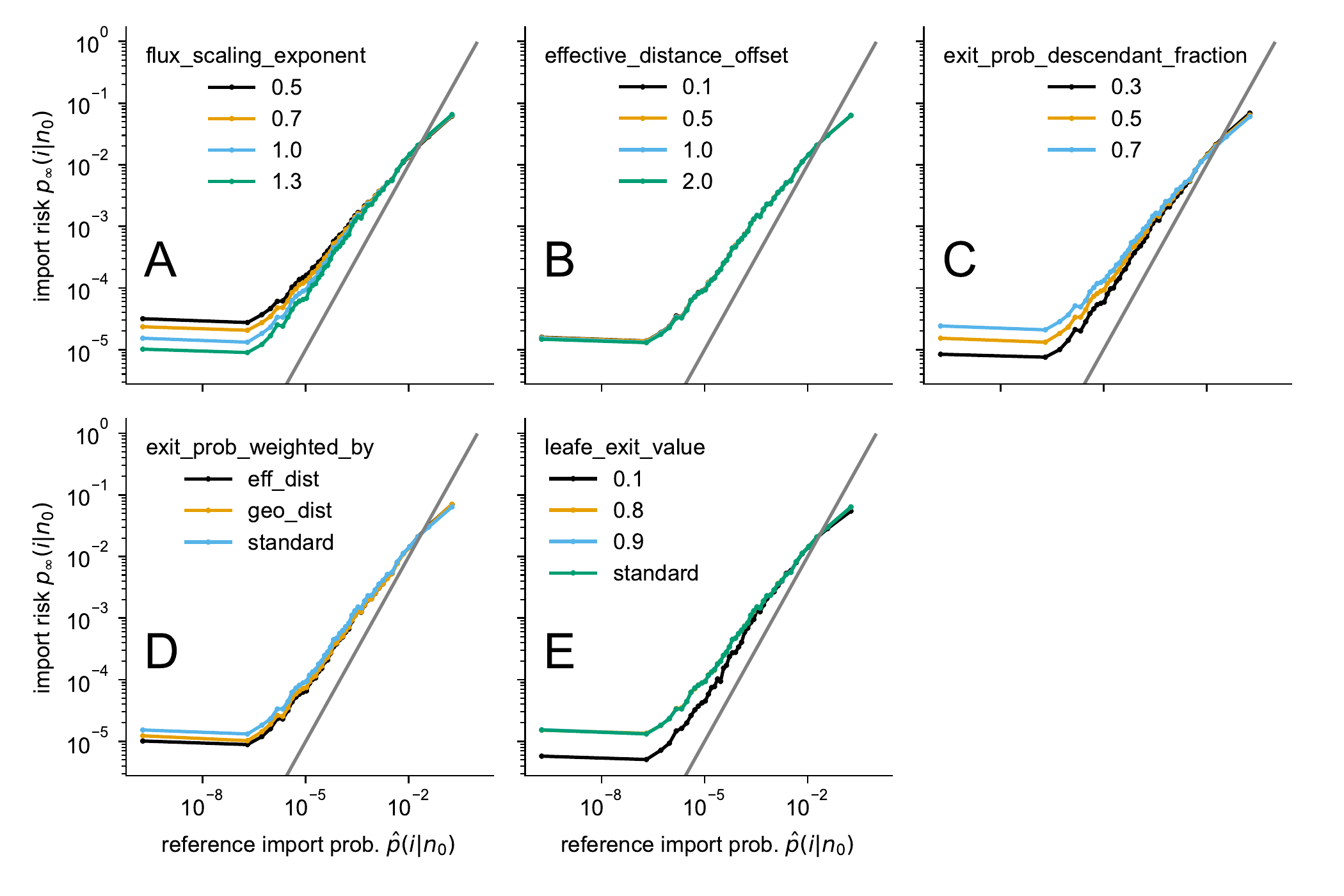}
    \caption{
        \textbf{Variations of the import risk model}
        to investigate how additional parameters influence the relation between the import risk and the reference import risk. 
        \textbf{A}: the flow scaling exponent $\nu$ that estimates the travelling population $N(i)$ of the airport $i$ depending on its WAN outflow $F_i$ via $N(i) = F_i^\nu$ (default: $\nu=1$).
        \textbf{B}: the effective distance offset $d_0$ that penalizes larger hop-distances in the effective distance $\deff(i|n_0)= d_0 - \ln(P_{ij})$ when creating the shortest path tree (default: $d_0=1$).
        \textbf{C}: the descendant fraction introduced in the shortest path exit probability, where $0.5$ is the default value and values larger than $0.5$ mean that the exiting at the descendant (or offspring) nodes compared to the current node becomes more likely.
        \textbf{D}: different weight options introduced for the shortest path tree exit probability. Per default, the node populations are not weighted. The weight is the inverse of either the geodesic or the effective distance.
        \textbf{E}: manually set shortest path exit probability of leaf nodes (dead-end nodes). Per default, the exit probability is $1$. A decrease to $0.9$ or $0.8$ does not visually change the median.
    }
    \label{fig:SI_import_risk_variations}
\end{figure}

We study the influence of the flow scaling exponent $\nu$ that estimates the travelling population $N_i$ of the airport $i$ depending on its WAN outflow $F_i$ via
\begin{align}
    N_i = F_i^\nu\ .
\end{align}
The larger the exponent, the smaller the difference to the reference data (Fig.~\ref{fig:SI_import_risk_variations}A).
That means if passengers are more likely to exit at larger airports, lower import probability are less strong overestimated. 

Instead, the effective distance offset $d_0$ does not change the overestimation at all (Fig.~\ref{fig:SI_import_risk_variations}B), which suggests that the differences in transition probability are too large to be influenced by penalizing hop distances.

Inspired by the fact that a larger exit at large airports decreases the overestimation and that large airports are rather at the beginning of the shortest path, we introduce a descendant fraction exit parameter $\mu$ that generalizes the shortest path exit probability from Eq.~\ref{eq:exit_prob} to 
\begin{align}
    q_i(n_0) = \frac{(1-\mu) N(i)}{ (1-\mu) N(i) + \mu N(\Omega(i|n_0))}\ .
\end{align}
With $\mu = 0.5$, we recover Eq.~\ref{eq:exit_prob} and with $\mu>0.5$ we shift the exit to the descendant (or offspring) nodes.
We verify the expected result, that an aversion of descendant exits decreases the overestimation of low import probabilities (Fig.~\ref{fig:SI_import_risk_variations}C).

Assuming that closer nodes on the shortest path tree are also geographically closer, we should observe a decrease in the overestimation if we weight the node populations with the inverse of their distance to the outbreak location (referred to in the main text as ``geodesic distance weighted'' exit probability).
We find the expected relation (Fig.~\ref{fig:SI_import_risk_variations}D), i.e. the distance weighted import risk model overestimates less.
Additionally, we find that overestimation further decreases if the effective distance is used for weighting.

Finally, we want to decrease the exit at nodes that have a large effective distance by setting the shortest path exit probability of the leaf- or dead-end nodes to a value smaller than the default value $1$.
The idea is that the random walker is not determined to end at leafs, but can walk on and is more likely return to hub nodes.
Interestingly, the expected decrease in overestimation is only present for low leaf exit values as $0.1$ (Fig.~\ref{fig:SI_import_risk_variations}E).

\begin{figure*}
    \centering
    \includegraphics[width=1\textwidth]{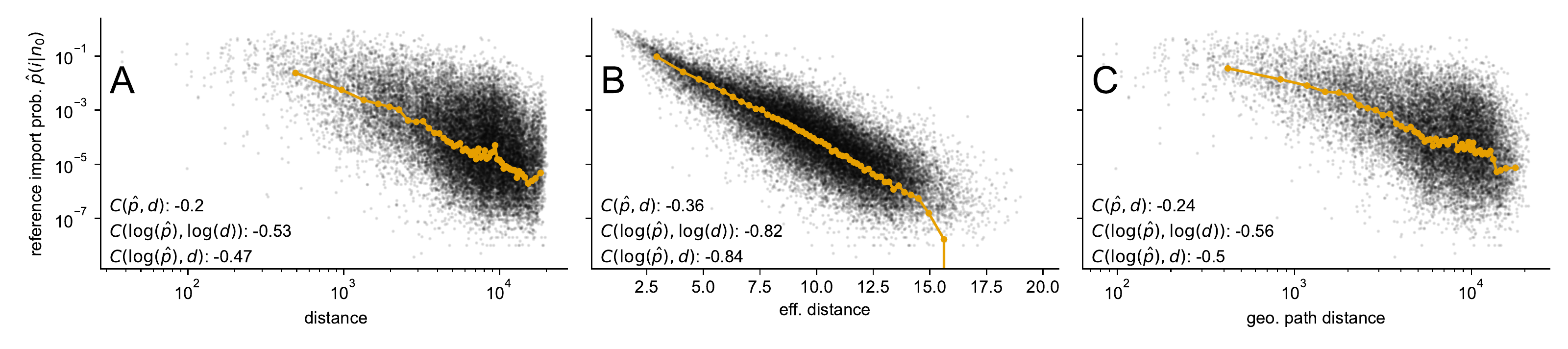}
    \caption{
        \textbf{Import probability dependence on}
        the geographic distance (\textbf{A}), the effective distance (\textbf{B}) and the geographic path distance (\textbf{C}).
        The orange line represents the median and $C(x, y)$ is the correlation between the two measures either log-transformed or not. 
        The geographic distance between countries is averaged over all airport pairs.
        The geographic path distance is the geographic distance along the shortest path derived from the WAN using $\deff$, i.e. it is a combination of geographic and network information.
        The axis scale corresponds to the one with the highest correlation, i.e. log-log for distance and path distance (\textbf{A}, \textbf{C}) and y-log for the effective distance (\textbf{B}).
    }
    \label{fig:ref_imp_vs_dists}
\end{figure*}

\begin{SCfigure*}
    \centering
    \includegraphics[width=0.6\linewidth]{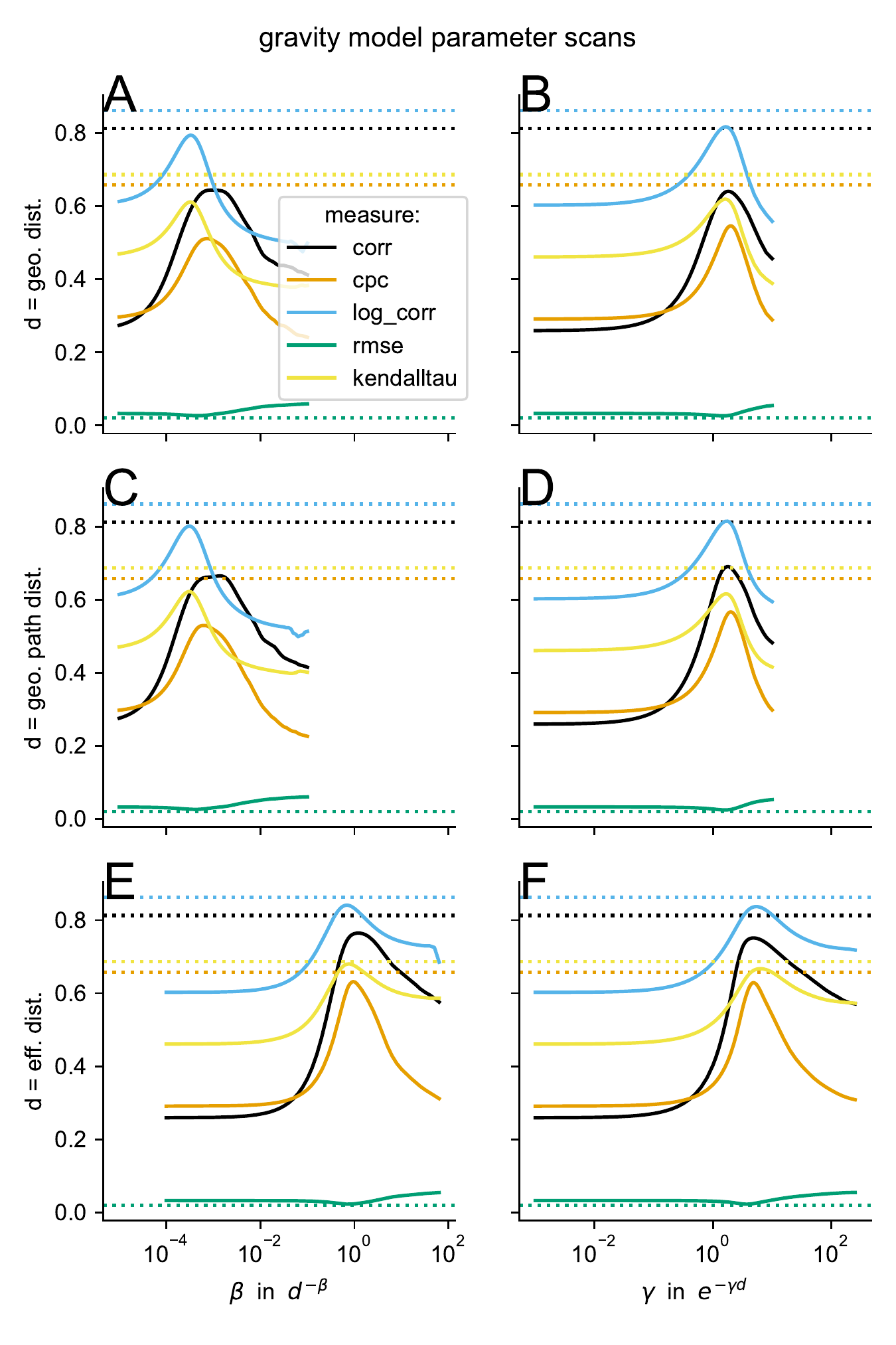}
    \caption{
        \textbf{Gravity model scans.}
        Parameter dependence of measures that compare the model estimated import probability with the reference import risk $\hat{p}(i|n_0)$.
        Thereby is "corr" the correlation, "cpc" the common part of commuters, "log\_corr" the correlation on log-scale, "rmse" the root mean squared error and "kendalltau" the rank correlation via Kendalls tau.
        Two versions of the gravity model are shown with 
        an exponentially decaying distance function $f(d)=e^{-\gamma d}$ (left column: \textbf{A, C, E}),
        and a power law decaying distance function $f(d)=d^{-\beta}$ (right column: \textbf{B, D, F}).
        As distance
        the geodesic distance (first row: \textbf{A, B}),
        the geodesic path distance (second row: \textbf{C, D}) and
        the effective distance (third row: \textbf{E, F}) are used.
        The dotted horizontal lines show the comparison measure with the import risk as model and have the same respective color.
        }
    \label{fig:SI_gravity_scans}
\end{SCfigure*}

\begin{SCfigure*}
    \centering
    \includegraphics[width=0.5\linewidth]{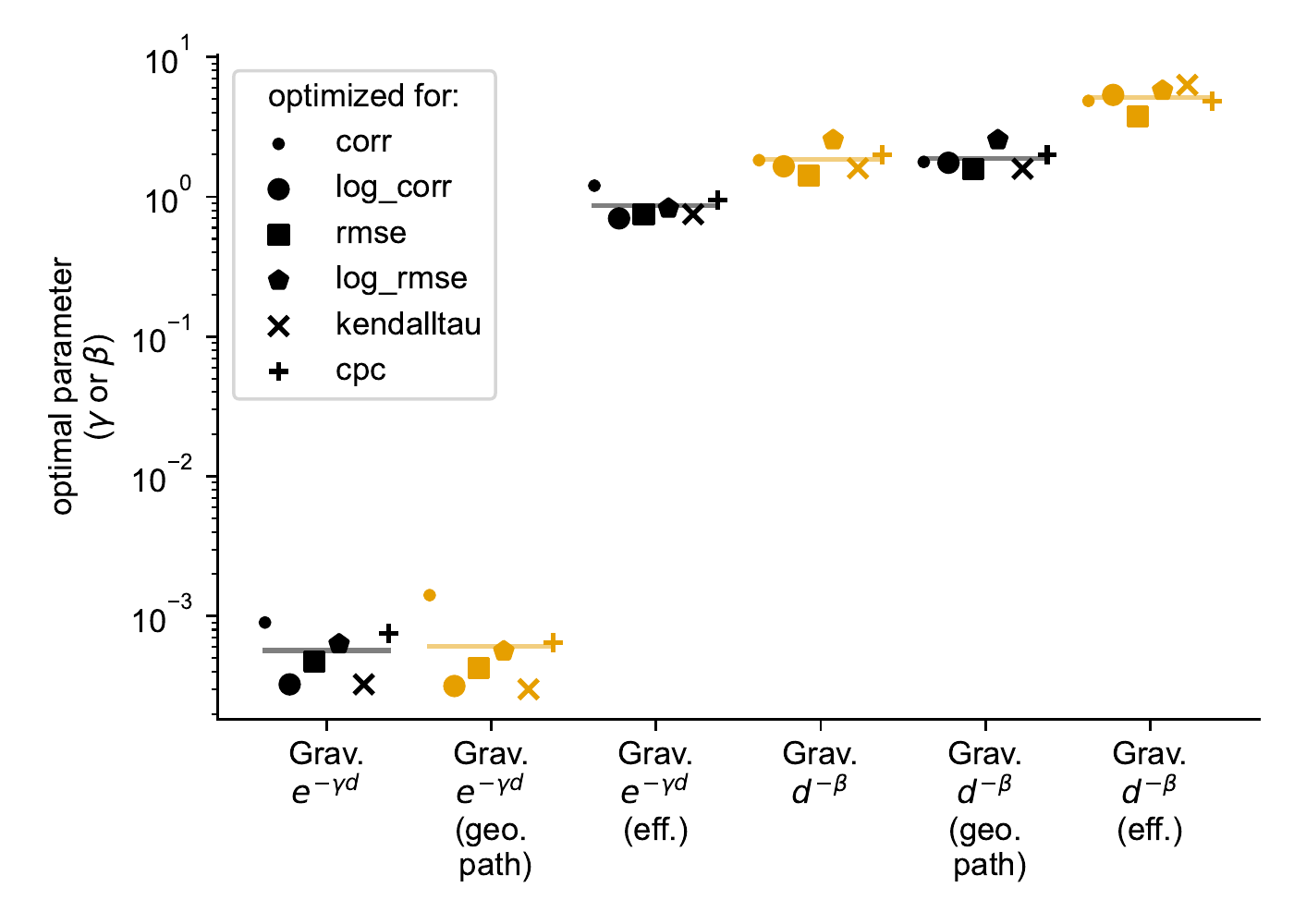}
    \caption{
        \textbf{Mean optimal parameters for gravity models.}
        For each gravity model with exponentially and power law decaying distance function and with one of the three different distance measures (geodesic distance, geodesic path distance and effective distance), the exponent $\gamma$ or $\beta$ that results in the best fit to the reference import risk is shown.
        The comparison is quantified via the correlation (corr), correlation between the log-transformed import risks (log\_corr), root mean square error (rmse), root mean square error of the log-transformed import risks (log\_rmse), Kendall rank correlation (kendalltau) and the common part of commuters (cpc).
        The mean optimal parameter for each model is marked by a horizontal line and their values are $\gamma = [5.68, 6.11]*10^{-4}$ for geographic and geo. path distance and $\gamma = 0.863$ for the effective distance, and $\beta=[1.84, 1,87, 5.41]$ for geo., geo. path, and effective distance, respectively.
    }
    \label{fig:SI_best_para_gravity}
\end{SCfigure*}

\begin{figure}
    \centering
    \includegraphics[width=1\linewidth]{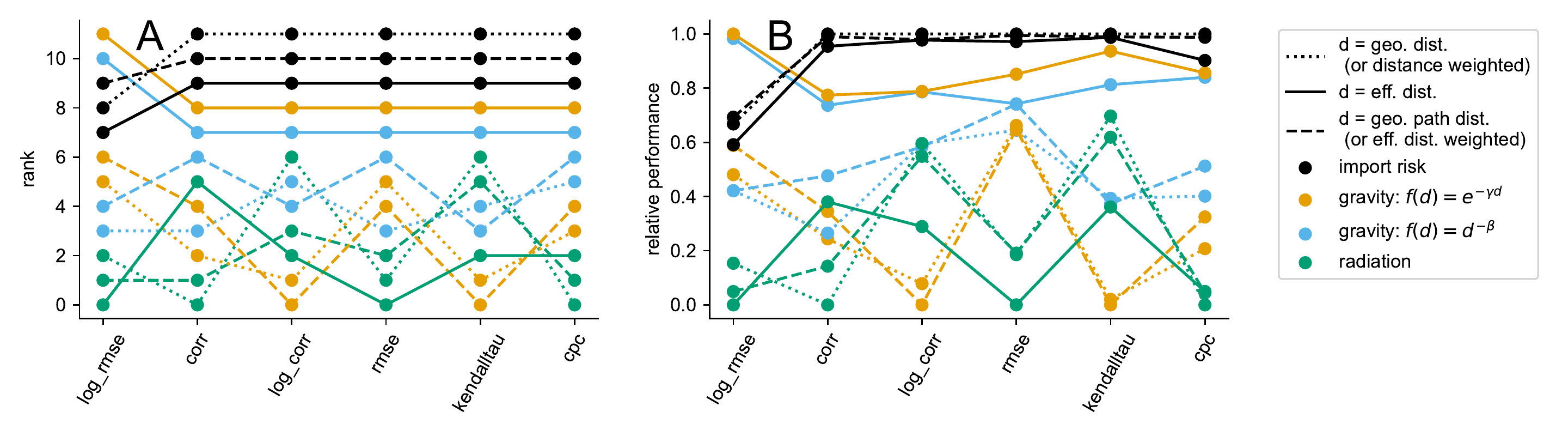}
    \caption{
        \textbf{Relative comparison measures for the import probability estimates.}
        The rank (\textbf{A}) and the relative performance (\textbf{B}) for the different import probability estimation models.
        The model that agrees best (worst) with the reference import risk according a specific measure has the highest (lowest) rank and a relative performance of one (zero).
        The relative performance is then a linear interpolation between the best and worst model.
        The comparison measures are the correlation (corr), correlation between the log-transformed import risks (log\_corr), Root-mean-square error (rmse), Root-mean-square error of the log-transformed import risks (log\_rmse), Kendall rank correlation (kendalltau) and the common part of commuters (cpc).
        As exponents of the gravity models the mean optimal parameter is used (horizontal lines in Fig.~\ref{fig:SI_best_para_gravity}).
    }
    \label{fig:SI_model_compare_lines}
\end{figure}

\begin{figure}
    \centering
    \includegraphics[width=1\linewidth]{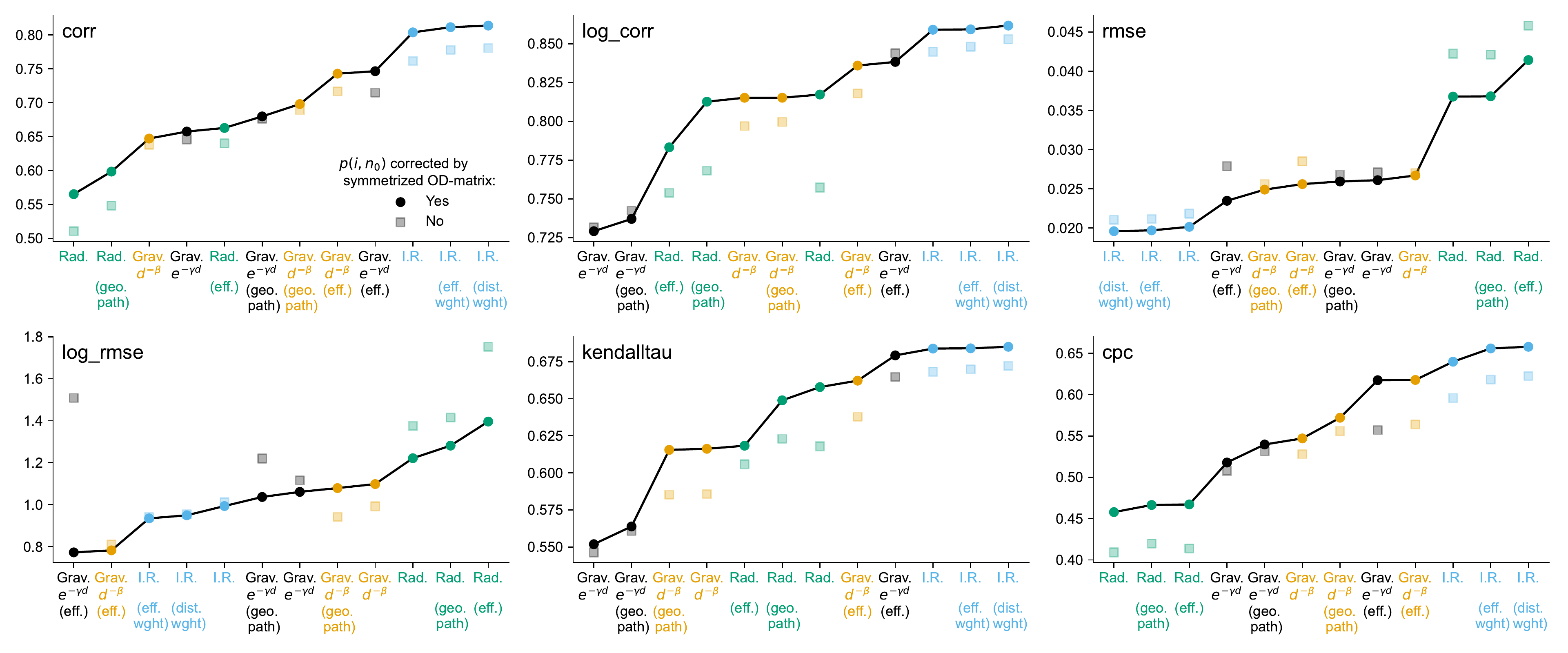}
    \caption{
        \textbf{Absolute comparison measures for the import probability estimates.}
        The comparison measures are the correlation (corr), correlation between the log-transformed import risks (log\_corr), Root-mean-square error (rmse), Root-mean-square error of the log-transformed import risks (log\_rmse), Kendall rank correlation (kendalltau) and the common part of commuters (cpc).
        As exponents of the gravity models the mean optimal parameter is used (horizontal lines in Fig.~\ref{fig:SI_best_para_gravity}).
        The colors depict the 4 different models.
        The solid circles are the models with corrected import probability by symmetrizing their OD-matrix, and the transparent squares are the non-corrected import probabilities of the respective model.
    }
    \label{fig:SI_model_comp_absolute}
\end{figure}

\begin{figure*}
    \centering
    \includegraphics[width=0.8\linewidth]{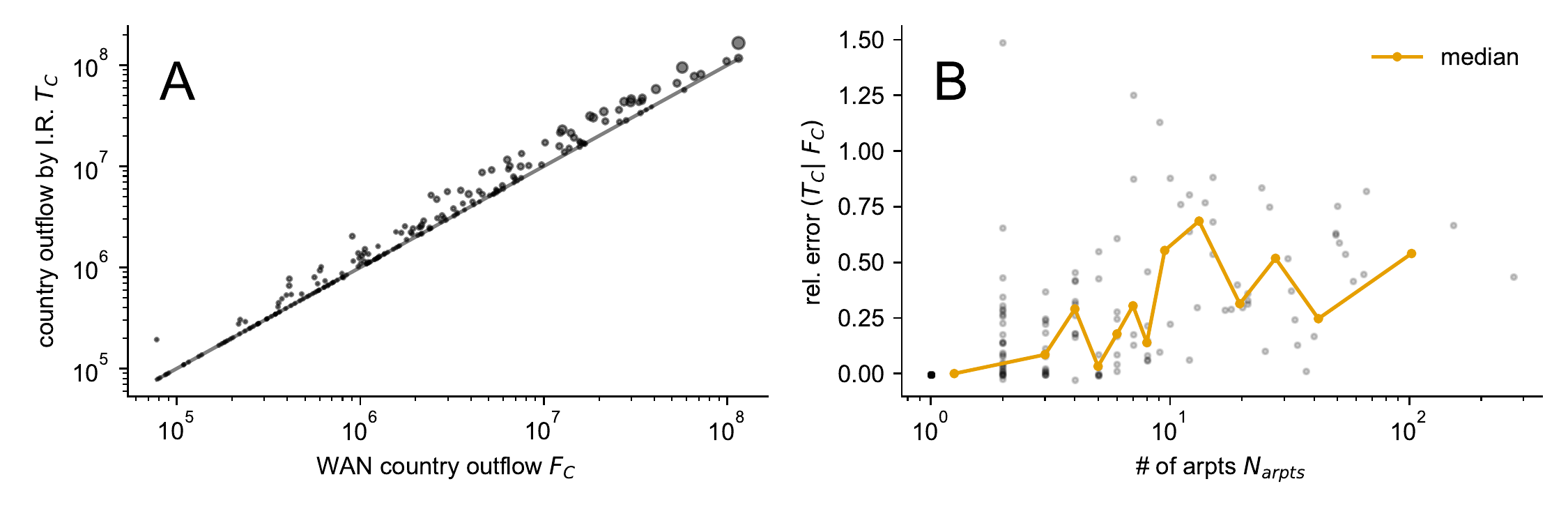}
    \caption{
        \textbf{Country outflow reconstruction by import risk.}
        The flow in the WAN leaving a country $F_C$ is estimated by the import risk model by $T_C = \sum_{n \in C} \sum_{m \not\in C}p_\infty(m|n) N_n$.
        Both measures are directly compared (\textbf{A}) and the relative error is computed depending on the number of airports in the respective country $N_{arpts}$ (\textbf{B}).
        The import risk model does not include the concept of a country which partly explains the overestimation for larger airports.
        Another explanation is the overestimation of the respective airport population $N_n=F_n$ by the WAN outflow for the import risk model (the true population is smaller because of the transit passengers that need to be excluded).
        Note that the WAN is used here, i.e. we check for self-consistency of the model and no reference data is included.
    }
    \label{fig:SI_self_consistent_outflow}
\end{figure*}

\begin{figure}
    \centering
    \includegraphics[width=1\textwidth]{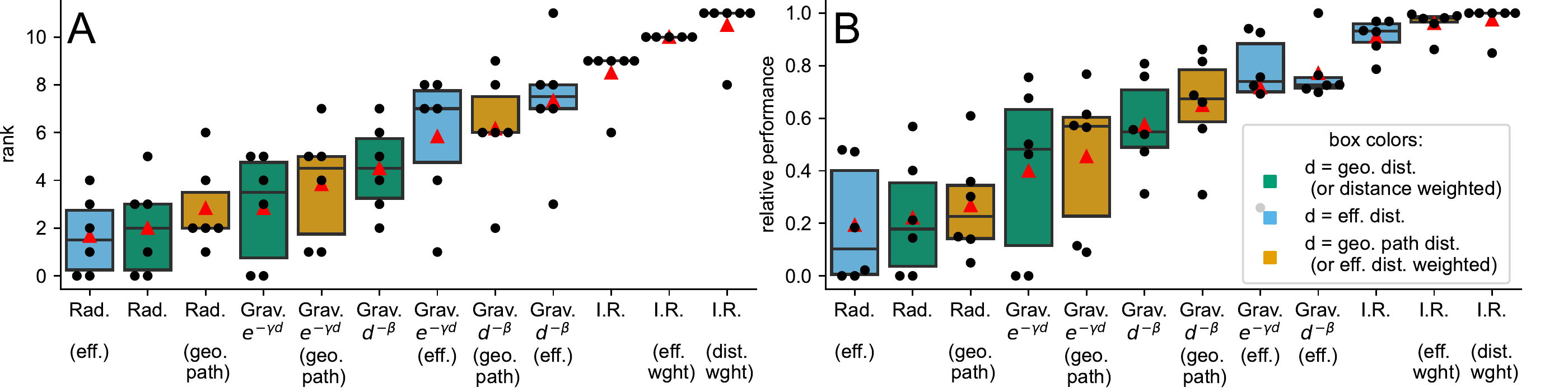}
    \caption{
        \textbf{Uncorrected models: rank and relative performance.}
        Same analysis as in the main text in Fig.~\ref{fig:rank_relperf}), however, here the uncorrected model predictions are used, i.e. without symmetrizing the OD-matrix.
    }
    \label{fig:SI_not_symmetrized_compare}
\end{figure}

\begin{figure}
    \centering
    \includegraphics[width=1\linewidth]{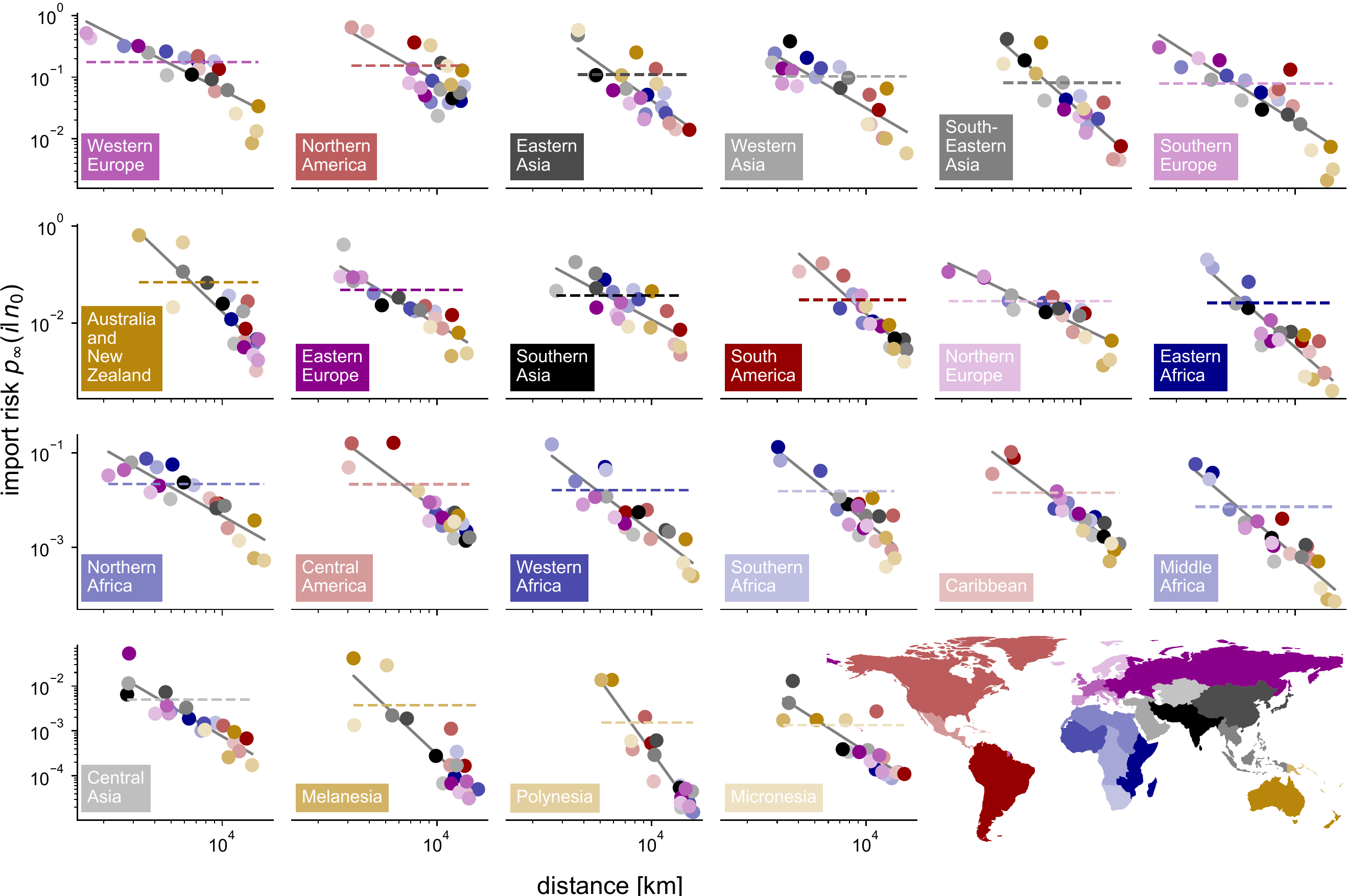}
    \caption{
        \textbf{Import risk between world regions to a specific target region.}
        In contrast to its derivation the import risk is displayed in a target-centric view, i.e. each panel displays the import probability to a single target region from all source regions.
        The distance between world regions is the mean distance between their airport locations.
        The grey line represents a power-law fit $p_\infty = c\cdot d^{-\alpha}$.
        The mean import risk is marked for each world region by a horizontal dashed line.
        The 22 target-world-regions are sorted according to their mean import risk.
    }
    \label{fig:SI_ir_dist_dep_all}
\end{figure}

\begin{figure}
    \centering
    \includegraphics[width=0.9\linewidth]{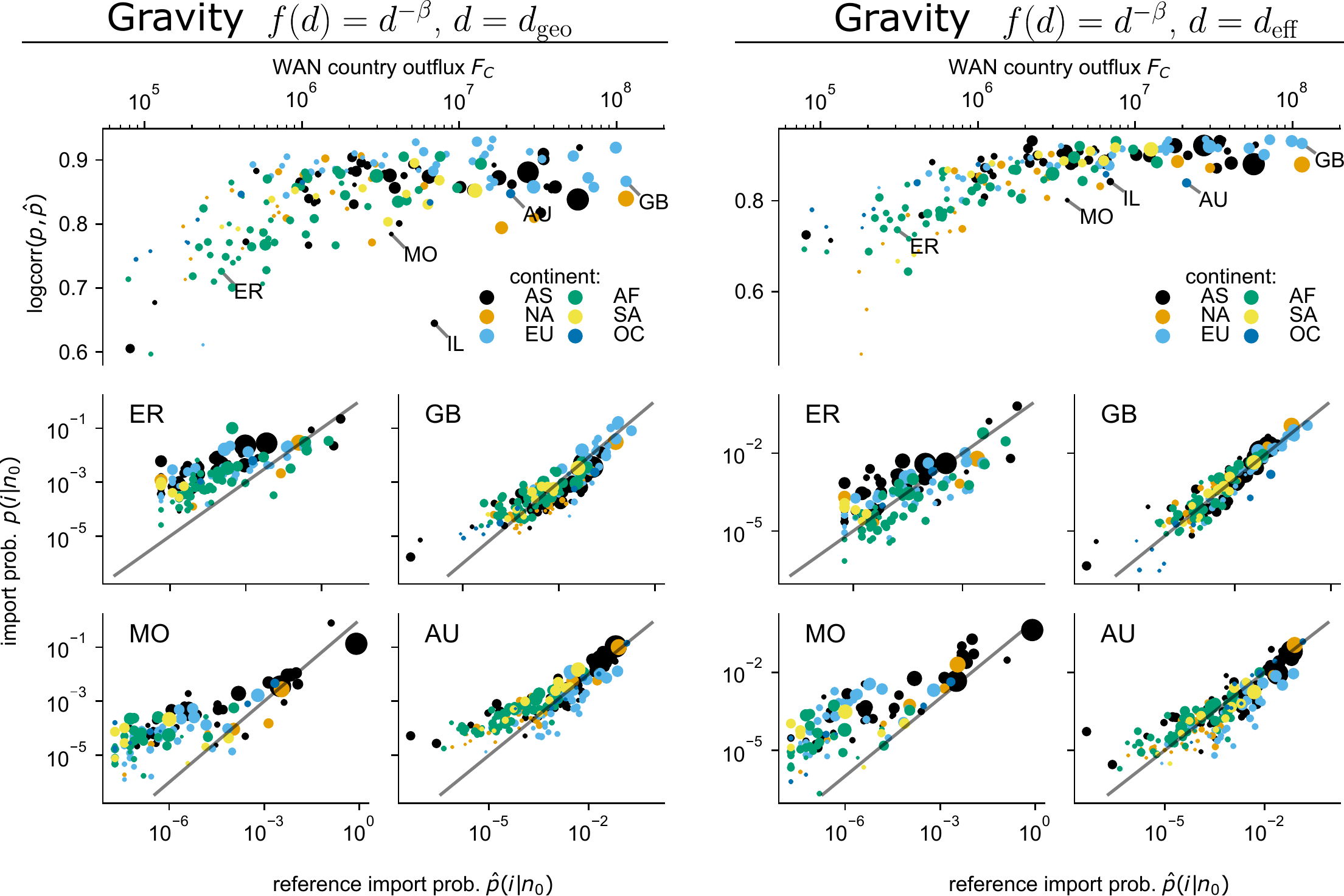}
    \caption{
        \textbf{Source countries prediction quality and WAN outflow for two gravity models.}
        Same model-result representation as in Fig.~\ref{fig:log_corr_and_examples} but here instead of the import risk model, the gravity model with power-law distance decaying function using the geodesic $d_{\mathrm{geo}}$ (left) or effective $\deff$ (right) distance is applied.
        Also for these models the $\logcorr$ between import probability estimates $p(i|n_0)$ and the reference data $\hat{p}(i|n_0)$ improves for countries with a larger outflow in the WAN.
    }
    \label{fig:SI_log_corr_and_examples}
\end{figure}

\begin{figure}
    \centering
    \includegraphics[width=0.8\linewidth]{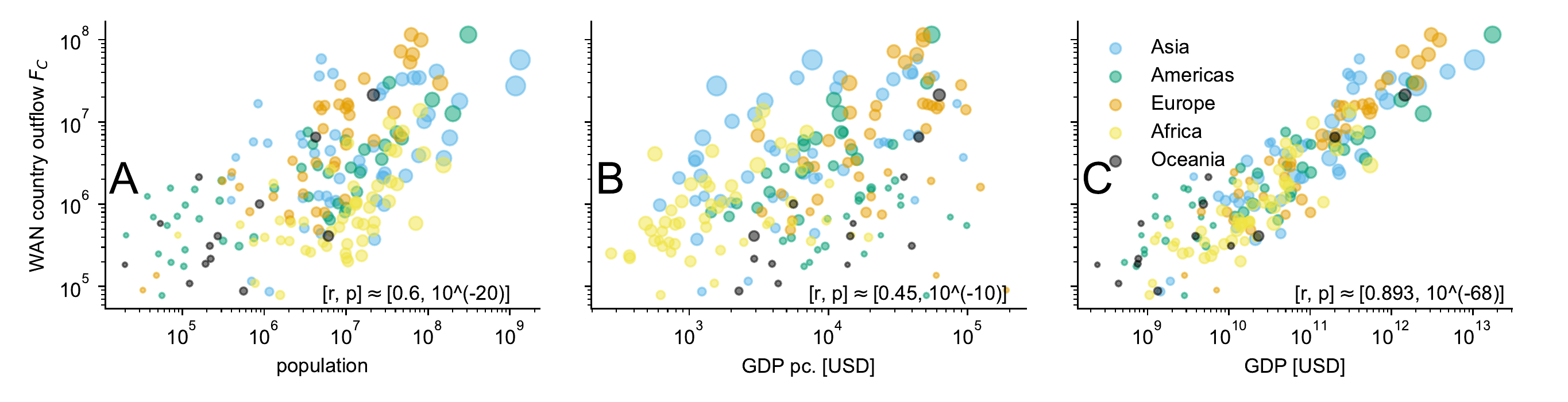}
    \caption{
        \textbf{WAN flow out of countries vs. population and GDP}
        The WAN flow out of a country is best mapped by its gross domestic product (GDP, \textbf{C}) compared to its population (\textbf{A}) or per capita GPD (\textbf{B}).
        The linear double-logarithmic regression results are shown in the lower part of each panel (r- and p-value).
        The size of each country corresponds to its population (\textbf{A}) and the color codes its continent.
        GDP is taken from the World Bank Dataset for the year 2014  \cite{WorldBank}.
    }
    \label{fig:SI_outflow_gdp}
\end{figure}

%TC:endignore
%the command above ignores the precedent section for word count

\end{document}